\documentclass{aastex631}
\usepackage[version=4]{mhchem}

\usepackage{booktabs, multirow}

\begin{document}

\title[lethal ozone concentrations on exoplanets]{Lethal surface ozone concentrations are possible on habitable zone exoplanets}

\author[0000-0001-6067-0979]{G. J. Cooke}\thanks{E-mail: gjc53@cam.ac.uk},
\affiliation{Institute of Astronomy, University of Cambridge, UK.\\}
\affiliation{School of Physics and Astronomy, University of Leeds, Leeds, LS2 9JT, UK\\}
\author[0000-0001-6699-494X]{D. R. Marsh}
\affiliation{School of Physics and Astronomy, University of Leeds, Leeds, LS2 9JT, UK\\}
\author[0000-0001-6078-786X]{C. Walsh}
\affiliation{School of Physics and Astronomy, University of Leeds, Leeds, LS2 9JT, UK\\}
\author[0000-0003-0304-7931]{F. Sainsbury-Martinez}
\affiliation{School of Physics and Astronomy, University of Leeds, Leeds, LS2 9JT, UK\\}

\begin{abstract}

\noindent Ozone (\ce{O3}) is important for the survival of life on Earth because it shields the surface from ionising ultraviolet (UV) radiation. However, the existence of \ce{O3} in Earth's atmosphere is not always beneficial. Resulting from anthropogenic activity, \ce{O3} exists as a biologically harmful pollutant at the surface when it forms in the presence of sunlight and other pollutants. As a strong oxidiser, \ce{O3} can be lethal to several different organisms; thus, when assessing the potential habitability of an exoplanet, a key part is determining whether toxic gases could be present at its surface. Using the Whole Atmosphere Community Climate Model version 6 (WACCM6; a three-dimensional chemistry-climate model), twelve atmospheric simulations of the terrestrial exoplanet TRAPPIST-1 e are performed with a variety of \ce{O2} concentrations and assuming two different stellar spectra proposed in the literature. Four atmospheric simulations of the exoplanet Proxima Centauri b are also included. Some scenarios for both exoplanets exhibit time-averaged surface \ce{O3} mixing ratios exceeding harmful levels of 40 ppbv, with 2200 ppbv the maximum concentration found in the cases simulated. These concentrations are toxic and can be fatal to most life on Earth. In other scenarios \ce{O3} remains under harmful limits over a significant fraction of the surface, despite there being present regions which may prove inhospitable. In the case that \ce{O3} is detected in a terrestrial exoplanet's atmosphere, determining the surface concentration is an important step when evaluating a planet's habitability.

\end{abstract}

\keywords{Exoplanets (498) --- Exoplanet atmospheres (487) --- Habitable planets (695) --- Exoplanet atmospheric composition (2021)}

\section{Introduction}
\label{Introduction section}

Molecular oxygen (\ce{O2}) makes up 21\% by volume of Earth's atmospheric composition and is required for aerobic respiration, provides a fuel for combustion, and gives rise to the `ozone layer'. In an atmosphere with \ce{O2} and sufficient ultraviolet (UV) irradiation at wavelengths shortward of 242.4 nm, \ce{O2} is photodissociated into atomic oxygen (O): 

\begin{align}
     \ce{O2 +} h\nu\ (175.9\ \textrm{nm} < \lambda < 242.4\ \textrm{nm}) &\ce{-> O + O}\label{O2 photolysis reaction 1},\\
      \ce{O2 +} h\nu\ (\lambda < 175.9\ \textrm{nm} ) &\ce{-> O(^1D) + O}, \label{O2 photolysis reaction 2}
\end{align}

\noindent where $h\nu$ represents a photon of frequency $\nu$, and $h$ is Planck's constant\footnote{\ce{O(^1D)} is the first excited state of atomic oxygen, where 1 represents the spin multiplicity and D is the spectroscopic notation for total orbital angular momentum. Note that the ground state of atomic oxygen can be written as \ce{O(^3P)}}. \ce{O2} and O can form ozone (\ce{O3}) via the following 3 body reaction:

\begin{align}
\ce{O + O2 + M &-> O3 + M}, \label{Ozone production reaction}
\end{align}

\noindent where \ce{M} is any third body (usually \ce{N2} or \ce{O2} on Earth due to their relatively high abundance). \ce{O3} can also be destroyed through photolysis, or by reacting with atomic oxygen:

\begin{align} 
    \ce{O + O3 &-> 2 O2}\label{ozone loss},\\
    \ce{O3 +} h\nu\ (\lambda \geq 320\ \textrm{nm}) &\ce{-> O2(^3\Sigma^-_g) + O},\label{ozone photolysis 1} \\
    \ce{O3 +} h\nu\ (\lambda \leq 320\ \textrm{nm}) &\ce{-> O2(^1\Delta_g) + O(^1D)}.\label{ozone photolysis 2}
\end{align}

On modern-day Earth, the majority of \ce{O3} resides in the stratosphere, roughly 15 to 35 km above the surface in the `ozone layer' \citep{2005ama..book.....B}. Here, \ce{O3} is beneficial for surface-dwelling life, absorbing biologically harmful UV radiation and providing a partial screen for life exposed to the Sun's radiation. Even though the majority of Earth's \ce{O3} is produced in the equatorial stratosphere, there exists a larger column of \ce{O3} at higher latitudes. This is because \ce{O3} is distributed through a seasonal equator-to-pole circulation driven by atmospheric gravity waves, known as the Brewer-Dobson circulation \citep{butchart2014brewer}. The Brewer-Dobson circulation has been observed to accelerate and decelerate due to climate change \citep{2008JAtS...65.2731G, butchart2014brewer, 2019ERL....14k4026F}, consequently affecting regional composition and temperatures near the tropopause and lower stratosphere.

The situation in the troposphere (the lowest atmospheric layer where temperature decreases with altitude) is rather different, because the photolysis rate of \ce{O2} is significantly lower here than in the stratosphere. Near the surface, volatile organic compounds (VOCs) can also contribute to \ce{O3} formation. Hydrocarbon emissions emanate from plants \citep[e.g., isoprene, $\alpha$-pinene;][]{1988Sci...241.1473C, sharkey2008isoprene} and, on modern day Earth, from anthropogenic activity \citep[e.g., naphthalene, acetone, formaldehyde, and many others;][]{2000AtmEn..34.2063A}. When photooxidation of hydrocarbons occurs in presence of nitrogen oxides, \ce{O3} can eventually be produced through the `smog mechanism' \citep{haagen1952chemistry}. For example, \ce{NO2} can be produced when \ce{OH} reacts with a hydrocarbon \citep{sillman1999relation}, \ce{RH} (R is an organic group), producing \ce{RO2} and \ce{H2O}:

\begin{align}
    \ce{RH + OH  &-> RO2 + H2O}.\label{smog reaction 1}
\end{align}

\ce{RO2} leads to \ce{NO2} formation via

\begin{align}
    \ce{RO2 + NO  &-> RO + NO2}\label{smog reaction 2}
\end{align}

\noindent Then, \ce{NO2} is photolysed in the presence of UV light:

\begin{align}
    \ce{NO2 +} h\nu\ \ce{&-> O + NO}\ (\lambda < \textrm{400 nm})\label{smog reaction 3}.    
\end{align}

The O produced can lead to reaction \ref{Ozone production reaction}, making \ce{O3} at low altitudes. On Earth, surface \ce{O3} is low at night when it is removed through reaction with NO,

\begin{equation}
    \ce{NO + O3 -> NO2 + O2},\label{O3 NO reaction}
\end{equation}

\noindent and \ce{O3} increases during the day due to photochemistry \citep{sillman1999relation}. Reaction \ref{O3 NO reaction} is part of a catalytic cycle, where a catalyst (\ce{X}) leads to the destruction of \ce{O3}, but is ultimately not used up in the overall reaction, such that

\begin{equation}
    \begin{split}
        \ce{X + O3 &-> XO + O2},\\
        \ce{XO + O &-> X + O2}, \\
        \textrm{Overall:}\ \ce{O3 + O &-> 2 O2}\label{Catalytic cycles}.
    \end{split}
\end{equation}

At the surface, \ce{O3} is considered a pollutant because it causes oxidative stress to plants, insects, and animals, including humans \citep{2011AtmEn..45.2284A, silva2013global, valavanidis2013pulmonary, 2014ACP....14.1011S, 2022ScTEn.827o4342D}. Oxidative stress is a chemical imbalance between oxidants and reductants inside an organism that can lead to molecular and biological damage \citep{lykkesfeldt2007oxidants, sies2017oxidative}. It has been demonstrated in many scenarios that \ce{O3} is an antimicrobial agent, capable of microbial inactivation of fungi, viruses, and bacteria  \citep{kim1999application, guzel2004use, najafi2009efficacy, fontes2012effect, epelle2023ozone, rojas2011research}. For instance, the removal of microbiota was demonstrated using ozonation of the air \citep{epelle2022bacterial}, and aqueous \ce{O3} is effective at inactivating microorganisms \citep{premjit2022aqueous}. Additionally, \ce{O3} has been found to be toxic across a wide range of organisms, including guinea pigs, rats, mice \citep{giese1954effects, stokinger1965ozone}, terrestrial plants \citep{rich1964ozone, bytnerowicz1993detecting, sandermann1996ozone, rao2001physiology, 2023ER....236k6816R}, aquatic life \citep{jones2006toxicity}, protozoa \citep{erickson2006inactivation}, and algae \citep{hu2003characteristic, gonccalves2011ozone}.

It is useful to consider some quantities to illustrate ozone's danger to life. For example, 40 parts per billion by volume (ppbv) of \ce{O3} is defined by the World Health Organisation (WHO) as a critical limit above which crop yield and species biomass may be reduced \citep{world2000air}. The WHO stated that significant health effects were exhibited by humans at 80 ppbv \citep{world2000air}, with \ce{O3} damaging lung function at 100 ppbv for 1 -- 8 hours of exposure. Indeed, years of evidence has indicated that long-term exposure to \ce{O3} appears to be related to premature human deaths \citep{bell2006exposure, turner2016long, 2022Innov...300246S}. For instance, \ce{O3} was attributed to 6,000 premature deaths in the EU in 2013 \citep{nuvolone2018effects}, and a modelling study by \cite{malashock2022global} calculated global \ce{O3}-attributable mortality in 2019 was 423,100 deaths (95\% confidence interval of 223,200 -- 659,400). The majority (77\%) of these were estimated to have occurred in Asia, where ground level \ce{O3} concentrations were relatively high \citep{2022ERL....17e4023M}. Furthermore, \cite{feng2022ozone} estimated that in East Asia, the reduced crop yield to \ce{O3} pollution costs US\$63 billion annually. If \ce{O3} is detrimental to life on Earth, then the same could be possible for extraterrestrial life. Due to ozone's powerful oxidising capacity \citep{menzel1984ozone, 2007WASP..187..285I}, it is possible that its toxicity to life could be ubiquitous. It is highly reactive, ranking amongst the highest oxidisers\footnote{\ce{F2} is the strongest oxidiser with a standard electrode potential of 2.87 eV, whilst \ce{O3} usually ranks second and has a standard electrode potential of 2.075 eV \citep{kishimoto2022effect}. The standard electrode potential is ``The value of the standard emf of a cell in which molecular hydrogen under standard pressure is oxidized to solvated protons at the left-hand electrode." \citep{mcnaught1997compendium}}. \ce{O3}, when internal to an organism, causes oxidative stress by releasing reactive oxygen species, which can then cause damage to proteins, DNA, and ultimately result in genetic mutations and cell growth that potentially turns into cancer \citep{klaunig2010oxidative}.

\ce{O3}, its spatial distribution on Earth, and its impact on terrestrial organisms, has been well studied. Less explored have been the implications of \ce{O3} in exoplanet atmospheres. Hundreds of terrestrial exoplanets, rocky planets orbiting stars other than the Sun, have been detected in our galaxy. Many of these are in the purported habitable zone (HZ) around their host star \citep[the region in which liquid water could persist on the surface of a rocky exoplanet;][]{1993Icar..101..108K}, although the potential for exoplanets and exomoons to be habitable goes beyond the traditional terrestrial-like HZ \citep[see e.g., ][]{2019ApJ...884..138C, 2020A&A...636A..50T, 2021ApJ...918....1M}. If extraterrestrial life exists, then at some point in its evolution, it is possible that \ce{O2} could be biologically produced just as it is on Earth, although there are several situations where \ce{O2} could be abiotically produced in high quantities \citep{2002AsBio...2..153D, 2014ApJ...785L..20W, 2014ApJ...792...90D, 2015AAS...22540704L, 2018ApJ...862...92K}. These scenarios include major water loss from photolysis \citep{2014ApJ...785L..20W, 2015AAS...22540704L} and high rates of \ce{CO2} photodissociation \citep{2015ApJ...806..249G, 2015ApJ...812..137H, 2016ApJ...819L..13S}. Either way, \ce{O3} is a molecule of interest because its detection can indicate the presence of atmospheric \ce{O2} \citep{1993A&A...277..309L, 2022A&A...665A.156K}. Additionally, \ce{O3} has strong spectroscopic signatures in both direct imaging and transmission spectra observations at relatively small volume mixing ratios \cite[e.g., between 10\textsuperscript{-7} and 10\textsuperscript{-5};][]{2017AsBio..17..287R, 2018AsBio..18..663S, 2022A&A...665A.156K}. Due to this property, some work has shown that in particular scenarios, \ce{O3} may be easier to detect than \ce{O2} \citep{ 2017AsBio..17..287R, 2022A&A...665A.156K, 2023MNRAS.518..206C}. To date, \ce{O3} has not yet been detected in the atmosphere of a terrestrial exoplanet, so the only estimates of the full \ce{O3} spatial distribution on exoplanets arises from three-dimensional chemistry-climate simulations. 

Tidally locked exoplanets are exoplanets that have a rotational period equal to their orbital period ($P$), such that they rotate synchronously \citep{1997Icar..129..450J, 2011ApJ...738...71S, 2019AnRFM..51..275P}. \cite{2018MNRAS.473.4672C} simulated tidally locked terrestrial exoplanets with orbital periods of 1 -- 100 days, finding that their atmospheric circulation depends in part on rotation rate. For $P < 25$ days, it was established that stratospheric transport could occur from the pole to the equator (described as an `Anti-Brewer Dosbon circulation'), or vice versa, depending on stratospheric wind breaking and the location of the planetary-scale Rossby waves (e.g., tropical or extratropical). At rotational periods greater than 25 days, the results from \cite{2018MNRAS.473.4672C} showed that a thermally driven circulation between the dayside and nightside could widely distribute air parcels. \cite{2020MNRAS.492.1691Y} used the Unified Model (UM) to simulate Proxima Centauri b (assuming a terrestrial exoplanet with a 11.18 day rotation period) in a slab ocean aquaplanet configuration, and found that the nightside \ce{O3} lifetime is much higher that it is on the dayside. The same conclusion was reached by \cite{2016EP&S...68...96P}, who simulated a tidally locked Earth with a rotational period of 365 days (no Brewer-Dobson circulation was present on this simulated exoplanet). \cite{2019ApJ...886...16C} used WACCM4, and reported that the pole-to-equator transport predicted by \cite{2018MNRAS.473.4672C} was present in two of their chemistry-climate simulations for terrestrial exoplanets with periods of 4.11 days and 7.91 days, and total irradiation of 1.0 $S_0$ and 1.1 $S_0$, respectively. Recently, \cite{2023MNRAS.526..263B} used the UM and found that \ce{O3} is produced on the dayside and transported to the nightside, with downwelling motions causing \ce{O3} to move into the troposphere at the positions of the nightside gyres. The use of a slab ocean aquaplanet configuration results in highly symmetric winds and chemical transport. 

Only a few studies have commented upon surface \ce{O3} in paleo atmospheres and exoplanet atmospheres. \cite{2013AsBio..13..415G} used a one-dimensional radiative-convective-photochemical model to investigate the atmospheric properties of super Earths around M0 to M7 stars and with surface gravity of 1 $g$ and 3 $g$ (where $g = 9.81$ m s\textsuperscript{-2}). Whilst the smog mechanism was important for \ce{O3} production around later spectral types, the surface \ce{O3} concentrations did not exceed harmful levels. \cite{2006IJAsB...5..295G} used a box model and showed how the smog mechanism could produce ground-level \ce{O3} up to 3500 ppbv during the Proterozoic (2.4 -- 0.541 Gyr ago) on Earth at 1\% the present atmospheric level (PAL) of \ce{O2}. During the Proterozoic, \ce{O2} concentrations could have ranged between $10^{-5}$ -- 1 times the PAL of \ce{O2} \citep{2019MinDe..54..485L,2020SciA....6.1420C, 2020PreR..343j5722S, 2021AsBio..21..906L}.  

The study by \cite{2006IJAsB...5..295G} is the only example of a simulated atmosphere which differs to modern Earth where harmful levels of \ce{O3} have been discussed, although the narrative focused on how \ce{O3} would have shielded the early Earth from UV radiation. No previous work has discussed the hypothetical dangers from \ce{O3} for extraterrestrial life on exoplanets and also used a 3D chemistry-climate model which accounts for horizontal transport. This work presents simulations of the exoplanets TRAPPIST-1 e and Proxima Centauri b using WACCM6, a 3D chemistry-climate model. Both exoplanets are located in the supposed HZ of their host stars, and TRAPPIST-1 e is a target for JWST transmission spectra observations. Proxima Centauri b orbits the star Proxima Centauri \citep[M5.5V spctral type, with a stellar effective temperature of 2,992 K;][]{2021ApJ...918...40P}, which is the closest star to the Sun \citep[1.3 pc;][]{2016A&A...595A...2G}, making it an exciting target for future observations \citep{2023arXiv230900725F}. TRAPPIST-1 e is a roughly Earth-sized exoplanet orbiting in the HZ around its ultra cool M8V \citep[stellar effective temperature of 2,566 K;][]{2021PSJ.....2....1A} dwarf host star, TRAPPIST-1. Whilst faint, TRAPPIST-1 is relatively close at a distance of 12.4 pc (40.5 light years). As host of 6 other terrestrial exoplanets, the TRAPPIST-1 system is a prime target to test theories of planetary system formation and evolution, by confirming whether atmospheres exist on any of the exoplanets, and characterising their properties if they do. To date, analysis of observations of the exoplanetary thermal emission with JWST suggests that the two innermost exoplanets, TRAPPIST-1b and c, have either thin atmospheres, or no atmosphere at all \citep{2023arXiv230610150Z, 2023Natur.618...39G}. Assuming that Earth-like atmospheres exist on both TRAPPIST-1 e and Proxima Centauri b, we investigate the abundance and distribution of \ce{O3} concentrations in different simulated scenarios, and discuss the implications for the habitability of oxygenated worlds.

\section{Simulations}
\label{Simulations}

\begin{table}[b!]
\caption{The sixteen simulations used in this study are listed. Twelve for TRAPPIST-1 e: six with the P19 spectrum \citep{2019ApJ...871..235P} and six with the W21 spectrum \citep{2021ApJ...911...18W}. There are four simulations of Proxima Centauri b, where the MUSCLES spectrum of Proxima Centauri (see text for details) is used as stellar input. Each simulation started with the pre-industrial (PI) WACCM6 simulation composition. Each set of six TRAPPIST-1 e simulations includes three with the present atmospheric level of \ce{O2} (0.21 by volume, all denoted as ``PI"); one where the substellar point is placed over the Pacific Ocean, one where it is placed over Africa (SPL), and one where it is not tidally locked and the rotation rate is 1 Earth day (noTL). Then, the 10\% PAL, 1\% PAL, and 0.1\% PAL simulations have reduced \ce{O2} mixing ratios from the PI simulation by 10, 100, and 1000 times, respectively. Each of the TRAPPIST-1 e simulations receive a total instellation of 900 W m\textsuperscript{-2}, and the Proxima Centauri b simulations receive 884 W m\textsuperscript{-2} of irradiation. The Proxima Centauri b simulations include the PI, 10\% PAL, 1\% PAL, and 0.1\% PAL cases. The simulated radius and mass of TRAPPIST-1 e are 0.91 $R_\oplus$ and 0.772 $M_\oplus$, respectively. For Proxima Centauri b, the radius and mass are 1.05 $R_\oplus$ and 1.07 $M_\oplus$, respectively. The $\oplus$ subscript denotes values relative to the Earth. The orbital parameters assume zero eccentricity and $0^\circ$ obliquity, and the table lists the period $P$ and the longitude of the substellar point (SP) relative to Earth's coordinates (the latitude of the SP is always $0^\circ$).  Each simulation has been run out for at least 250 model Earth years.}
\centering
\begin{tabular}{@{}ccccc@{}}
\toprule
Simulation & Planet  & Spectrum & \ce{O2} mixing ratio [PAL] & Orbital parameters \\ \hline
W21 PI & TRAPPIST-1 e  & W21 & 1.000 & $P = 6.1$d, SP = $180^\circ$ lon \\
W21 PI noTL & TRAPPIST-1 e  & W21 & 1.000 & $P = 1$d \\
W21 PI SPL & TRAPPIST-1 e  & W21 & 1.000 & $P = 6.1$d,  SP = $30^\circ$ lon \\
W21 10\% PAL & TRAPPIST-1 e & W21 & 0.100 & $P = 6.1$d,  SP = $180^\circ$ lon \\
W21 1\% PAL & TRAPPIST-1 e  & W21 & 0.010 & $P = 6.1$d,  SP = $180^\circ$ lon \\
W21 0.1\% PAL & TRAPPIST-1 e & W21 & 0.001 & $P = 6.1$d,  SP = $180^\circ$ lon \\
P19 PI   & TRAPPIST-1 e     & P19 & 1.000 & $P = 6.1$d,  SP = $180^\circ$ lon \\
P19 PI noTL & TRAPPIST-1 e  & P19 & 1.000 & $P = 1$d \\
P19 PI SPL & TRAPPIST-1 e  & P19 & 1.000 & $P = 6.1$d, SP = $30^\circ$ lon \\
P19 10\% PAL & TRAPPIST-1 e & P19 & 0.100 & $P = 6.1$d,  SP = $180^\circ$ lon \\
P19 1\% PAL & TRAPPIST-1 e  & P19 & 0.010 & $P = 6.1$d,  SP = $180^\circ$ lon \\
P19 0.1\% PAL & TRAPPIST-1 e & P19 & 0.001 & $P = 6.1$d,  SP = $180^\circ$ lon \\
PCb PI & Proxima Centauri b & PC MUSCLES & 1.000 & $P = 11.18$d, SP = $180^\circ$ lon \\
PCb 10\% PAL & Proxima Centauri b & PC MUSCLES & 0.100 & $P = 11.18$d, SP = $180^\circ$ lon \\
PCb 1\% PAL & Proxima Centauri b & PC MUSCLES & 0.010 & $P = 11.18$d, SP = $180^\circ$ lon \\
PCb 0.1\% PAL & Proxima Centauri b & PC MUSCLES & 0.001 & $P = 11.18$d, SP = $180^\circ$ lon \\ \hline
\end{tabular}
\label{Simulation table}
\end{table}

\begin{figure*}[t!]
	\centering
	\includegraphics[width=1\textwidth]{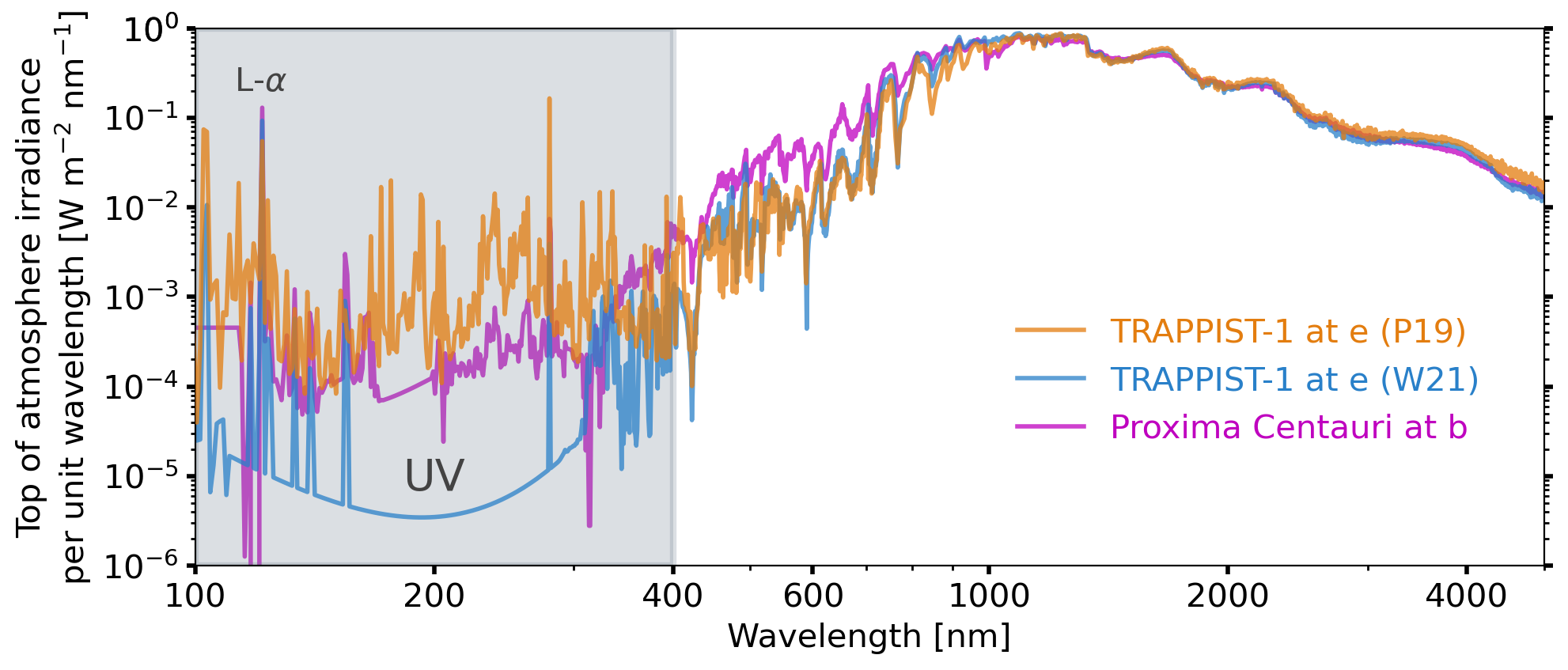}
    \caption{The three stellar input spectra used for the WACCM6 simulations are the PC MUSCLES spectrum for Proxima Centauri at b (magenta), the \cite{2019ApJ...871..235P} (P19; orange) and \cite{2021ApJ...911...18W} (W21; blue) spectra for TRAPPIST-1 at e. The top of atmosphere irradiance per unit wavelength is shown against the wavelength in nm. In the simulations, TRAPPIST-1 e receives $900~\textrm{W\,m}^{-2}$ of irradiation (0.66\ $S_\oplus$, where $S_\oplus$ is the total insolation received by the Earth), and Proxima Centauri b receives $884~\textrm{W\,m}^{-2}$ (0.65\ $S_\oplus$). The UV range is highlighted in grey between 100 -- 400 nm, and the Lyman-$\alpha$ line is labelled. The average difference between the TRAPPIST-1 spectra in the UV range is a factor of 500, with a difference of up to 5000 in some wavelength bins.} 
    \label{TP-1 spectra figure}
\end{figure*}

We use the three-dimensional Earth System Model WACCM6 \citep{2019JGRD..12412380G} to perform twelve simulations of possible climates of TRAPPIST-1 e, and four possible climate simulations of Proxima Centauri b. WACCM6 is a configuration of the Coupled Earth System Model (CESM), and we use version 2.1.3 (CESM2.1.3). In each simulation the initial conditions represent the approximate conditions of Earth's pre-industrial (PI) atmosphere for the year 1850. The simulations have the modern ocean and land configuration, a horizontal resolution of $1.875^\circ$ by $2.5^\circ$ (96 latitude points and 144 longitude points), and 70 vertical atmospheric levels from 1000 hPa to $4.5 \times 10$\textsuperscript{-6} hPa. Both the atmosphere and ocean models are set up to be fully interactive so that they respond to physical changes such as temperature. Because it is likely that Proxima Centauri b and TRAPPIST-1 e rotate synchronously \citep[they may be tidally locked to their host star, although spin-orbit resonance states are plausible;][]{2016A&A...596A.111R, 2021PSJ.....2....4R}, the substellar point is fixed. This is done by fixing the solar zenith angle in each grid cell. The exoplanet’s obliquity and orbital eccentricity are set to zero.  We run WACCM6 with middle atmosphere chemistry which is described in \cite{2020JAMES..1201882E}, where further details can be found. This chemical mechanism in the WACCM6 simulations has 98 chemical species and 298 chemical reactions, including both the photochemical and heterogeneous reactions that are necessary to simulate the atmospheric conditions of 1850, and crucially, \ce{O3} chemistry. \ce{O3} pollution due to VOCs is not simulated. The atmospheric time step, $\Delta t$, is 30 minutes. The concentrations of 75 species are computed using the implicit method, which considers the chemical system at time $t$ and $t+\Delta t$ to evaluate the system at the future time step $t+\Delta t$ \citep{sandu1997benchmarking}. 22 long-lived species are computed with the explicit method, which calculates the chemical system at a later time $t+\Delta t$ by considering the current system at time $t$ \citep{2005ama..book.....B}. \ce{N2} is invariant in each simulation, and its mixing ratio in each simulation is adapted to ensure that the atmosphere maintains a surface pressure of 1,000 hPa. Following the work by \cite{2023RSOS...1030056J}, we include absorption by \ce{O3}, \ce{O2}, \ce{CO2} and \ce{H2O} in the Schumann-Runge bands (175 -- 192 nm). 

`Dry deposition' is the process through which atmospheric trace gases and particulate matter are deposited on Earth's surface and are removed from the atmosphere, and is an atmospheric sink of \ce{O3}. Dry deposition in WACCM6 \citep{2020JAMES..1201882E} was updated following \cite{2014GeoRL..41.2988V}, and was originally based on a parameterisation from \cite{1989AtmEn..23.1293W}. The parameterisation accounts for variables such as the aerodynamic resistance and the surface resistance, and is influenced by vegetation, if present.  

We assume TRAPPIST-1 e receives $900\ \textrm{W m}^{-2}$ of irradiation and that Proxima Centauri b receives $884\ \textrm{W m}^{-2}$  (0.66 $S_0$ and 0.65 $S_0$ respectively, where $S_0$ is the total insolation that the Earth receives). This is consistent with previous work on Proxima Centauri b \citep{2017A&A...601A.120B,2020MNRAS.492.1691Y, 2022MNRAS.517.2383B, 2023MNRAS.518.2472R} and with the TRAPPIST-1 Habitable Atmosphere Intercomparison project \citep[THAI;][]{2020GMD....13..707F,2022PSJ.....3..211T}. Proxima Centauri b was detected using the radial velocity method and has a minimum mass measured of $M_P \sin{i} = 1.07\ M_\oplus$ \citep{2022A&A...658A.115F} only, where $M_P$ is the mass of the exoplanet and $i$ is the inclination angle of the planetary orbit. Therefore, a recently estimated mass-radius relationship from \cite{2020A&A...634A..43O}, given as $R_p = 1.03\ M_p^{0.29}$, was used to estimate the planetary radius. Assuming an optimistic mass of $M_P = 1.07\ M_\oplus$, this places the radius of Proxima Centauri as $1.05\ R_\oplus$, and the surface gravity of Proxima Centauri b at 12.2 m s\textsuperscript{-2}. In our simulations, TRAPPIST-1 e is set to have a mass of $0.772\ M_\oplus$ and a radius of $0.91\ R_\oplus$, consistent with the THAI project and transit timing variations (TTVs) from \cite{2018A&A...613A..68G}. The surface gravity of TRAPPIST-1 e is therefore set to 9.14 m s\textsuperscript{-2}. 

Two semi-empirical stellar spectra were used in the TRAPPIST-1 e simulations. \cite{2019ApJ...871..235P}, henceforth known as P19, modelled the stellar energy distribution (SED) of TRAPPIST-1 and 
produced model 1A, 2A, and 2B, of which we use model 1A \cite[version 1;][]{t9-j6bz-5g89}. More recently, \cite{2021ApJ...911...18W}, hereafter known as W21, used further HST observations to produce a semi-empirical SED of TRAPPIST-1 \citep[version 1;][]{zenodo.4556130} as part of the Mega-MUSCLES series \citep{2019ApJ...871L..26F, 2021ApJ...911...18W}. Details of their spectra can be found in the aforementioned references. Both the stellar spectra are included here to illustrate how different strengths and shapes of the incoming UV radiation environment can affect the abundance and distribution of surface \ce{O3}. For Proxima Centauri b, we use the GJ 551 MUSCLES \citep[version 2.2;][]{2016ApJ...820...89F, 2016ApJ...824..101Y, 2016ApJ...824..102L} spectrum as input\footnote{\href{https://archive.stsci.edu/prepds/muscles/}{GJ 551 found at https://archive.stsci.edu/prepds/muscles/}}. GJ 551 is the Gliese–Jahreiß catalog name for Proxima Centauri.

Note that the TRAPPIST-1 e simulations were started in the year 2020, before \cite{2021PSJ.....2....1A} published updates to planetary parameters in the TRAPPIST-1 system. For TRAPPIST-1 e, \cite{2021PSJ.....2....1A} gave mass and radius values of $0.69 M_\oplus$ and $0.92\ R_\oplus$, respectively, meaning the surface gravity would be 8.015 m s\textsuperscript{-2}, instead of 9.14 m s\textsuperscript{-2} as used here. Using these updated values, the scale height of the atmosphere would increase, but we expect that simulations with the parameters from \cite{2021PSJ.....2....1A} would produce similar surface \ce{O3} mixing ratios to the ones we present here. Only the minimum mass has been measured for Proxima Centauri b, so it is conceivable that it may have a larger mass and radius than the values used here. \cite{2016ApJ...831L..16B} estimated the radius to be in the range 0.94 -- $1.40 R_\oplus$, placing it somewhere between a Mercury-like exoplanet and an ocean-like world. Regardless, with M-dwarf stars being so numerous, it is plausible that somewhere there exits an exoplanet with similar size and instellation, such that these simulations remain useful should Proxima Centauri b eventually be confirmed to have a mass or radius which is significantly larger. 

A summary of the simulations is given in Table \ref{Simulation table}. For TRAPPIST-1 e, six simulations use the P19 (stronger UV) spectrum, and six simulations use the W21 (weaker UV) spectrum. For both exoplanets, atmospheric concentrations of \ce{O2} at the present atmospheric level (PI; 0.21 by volume), 10\% PAL, 1\% PAL, and 0.1\% PAL, are simulated. For TRAPPIST-1 e, we move the substellar point for the 100\% PAL simulation between the Pacific Ocean ($180^\circ$ longitude; PI case) and Africa ($30^\circ$ longitude; PI SPL case). We run two simulations which are not tidally locked and have a rotational period of 1 day. Whether a slab or dynamic ocean is implemented, and whether the land or ocean is at the substellar point, can modulate the climatology of exoplanets \citep{2014PNAS..111..629H, 2018ApJ...854..171L, 2019AsBio..19...99D, 2020ApJ...896L..16S, 2021ApJ...910L...8Z, 2022MNRAS.513.2761M, 2022GeoRL..4995748O}. \cite{2020ApJ...896L..16S} found that broad climate differences between models (ROCKE-3D and the UM) were larger than the difference between a slab ocean and a dynamic ocean. Because of this previous work, and the fact that the PI and PI SPL simulations show only small globally averaged chemical differences in \ce{O3}, and due to computational expense (WACCM6 takes 1,332 core-hours per simulated year to run), we do not simulate the substellar point over land in any of the reduced \ce{O2} cases. Each simulation has been run out for at least 250 model Earth years, and then we present the last year of data (365 Earth days). The full details of the model set-up, alongside simulation scripts, are available via GitHub\footnote{\href{https://github.com/exo-cesm/CESM2.1.3/tree/main/Tidally_locked_exoplanets}{https://github.com/exo-cesm/CESM2.1.3/tree/main/Tidally\textunderscore locked\textunderscore exoplanets}}.

\section{Results}
\label{Results section}

\begin{figure*}[t!]
	\centering
	\includegraphics[width=1\textwidth]{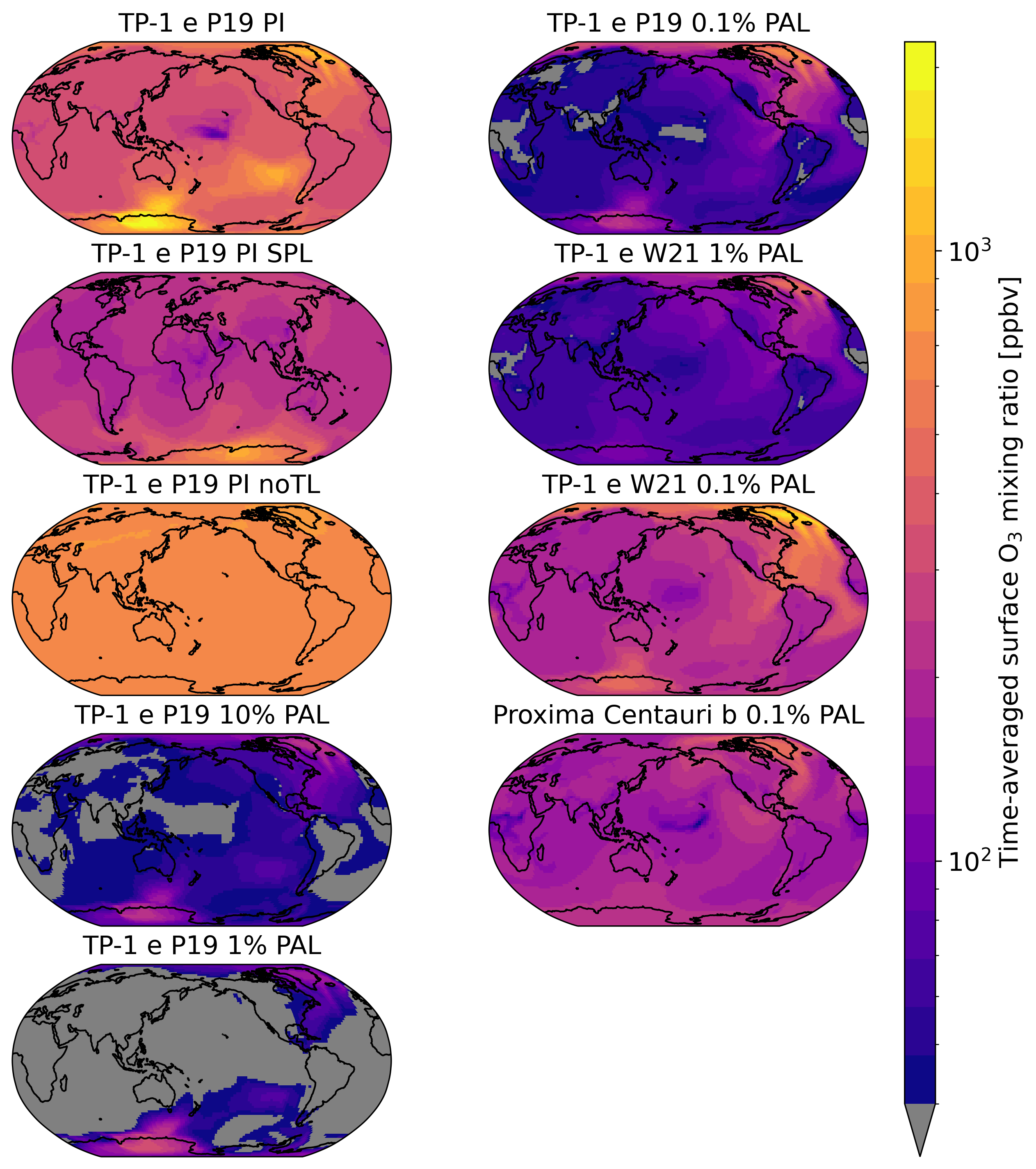}
    \caption{The surface mixing ratio of \ce{O3}, in parts per billion by volume (ppbv), is displayed for the simulations in this work which have time-averaged surface \ce{O3} mixing ratios exceeding 40 ppbv. This includes all TP-1 e P19 cases, the TP-1 e W21 1\% PAL and 0.1\% PAL simulations, and only a single Proxima Cenaturi b case (0.1\% PAL). The PI cases start with an initial pre-industrial atmospheric composition. The PI case has the substellar point placed over the Pacific Ocean, the SPL case has it placed over Africa, and the noTL case is not tidally locked, so that the substellar point moves with time. PAL is the present atmospheric level of \ce{O2}, which is a mixing ratio of 21\% by volume. The TP-1 e P19 simulations have stronger incident ultraviolet radiation than the TP-1 e W21 simulations. See Table \ref{Simulation table} for a more detailed description of the simulations. Grey indicates where the \ce{O3} mixing ratio is below 40 ppbv and thus at ``safe" levels, whilst the different shades of yellow-orange-purple indicate places that exceed 40 ppbv, i.e., these concentrations are known to be harmful to life on Earth. The colour map has a log scale which extends from 40 ppbv to 2200 ppbv.} 
    \label{Surface ozone figure - TP-1 e}
\end{figure*}

\subsection{Surface ozone concentrations}

Fig. \ref{Surface ozone figure - TP-1 e} shows the time-averaged distribution of \ce{O3} at the atmospheric level closest to surface for the cases which have time-averaged surface \ce{O3} mixing ratios of 40 ppbv or greater. We use 40 ppbv as a lower cut-off for harmful levels of surface \ce{O3} \citep{world2000air}. Grey indicates regions below 40 ppbv, whilst the yellow-orange-purple colour map indicates regions where \ce{O3} exceeds 40 ppbv. 

The TRAPPIST-1 e (TP-1 e) P19 PI, PI SPL, no TL, and W21 0.1\% PAL simulations everywhere exceed 40 ppbv for surface \ce{O3}. The P19 PI simulation has a maximum mixing ratio of 2200 ppbv, which is the largest surface mixing ratio in all of the simulations presented. In the P19 PI and 0.1\% PAL simulations and the W21 0.1\% PAL simulation, specific locations (e.g., Antarctica or Greenland) have extremely high mixing ratios, exceeding 1000 ppbv, which is deadly to some organisms on Earth. On the other hand, the W21 PI, PI SPL, and noTL simulations everywhere have \ce{O3} mixing ratios below 40 ppbv, and are not shown. The low \ce{O3} surface concentrations are a consequence of the upper atmosphere efficiently absorbing UV such that insufficient UV reaches altitudes closer to the planetary surface to synthesise enough \ce{O3}. The P19 10\%, 1\%, and 0.1\% PAL simulations, and the W21 1\%, simulation, have some areas where \ce{O3} exceeds 40 ppbv, whilst maintaining regions below this limit. For the Proxima Centauri b (PCb) cases, the PI, 10\% PAL, and 1\% PAL cases have surface \ce{O3} levels below 40 ppbv everywhere. For the 0.1\% PAL PCb scenario, surface \ce{O3} everywhere exceeds 40 ppbv and has an global mean mixing ratio of 203 ppbv.

In terms of time variability, the surface \ce{O3} concentrations are not static. Taking the last year of simulated data and averaging each calendar month, the fraction of land for each simulation where surface \ce{O3} concentrations are under 40 ppbv is given in Table \ref{Dry dep table}. For example, the P19 10\%, 1\%, and 0.1\% PAL simulations have monthly \ce{O3} surface concentrations under harmful levels varying between 12 -- 44\%, 75 -- 83\%, and 4 -- 9\% of the total surface area. Considering all these scenarios, the prospect is raised for safe areas on exoplanets which are sheltered from hazardous \ce{O3} concentrations found at other locations. Meanwhile, some locations will fluctuate between toxic and safe levels. Only the P19 PI noTL simulation has surface \ce{O3} mixing ratios that everywhere exceed 40 ppbv throughout the final year of data.

\begin{figure*}[t!]
	\centering
	\includegraphics[width=1\textwidth]{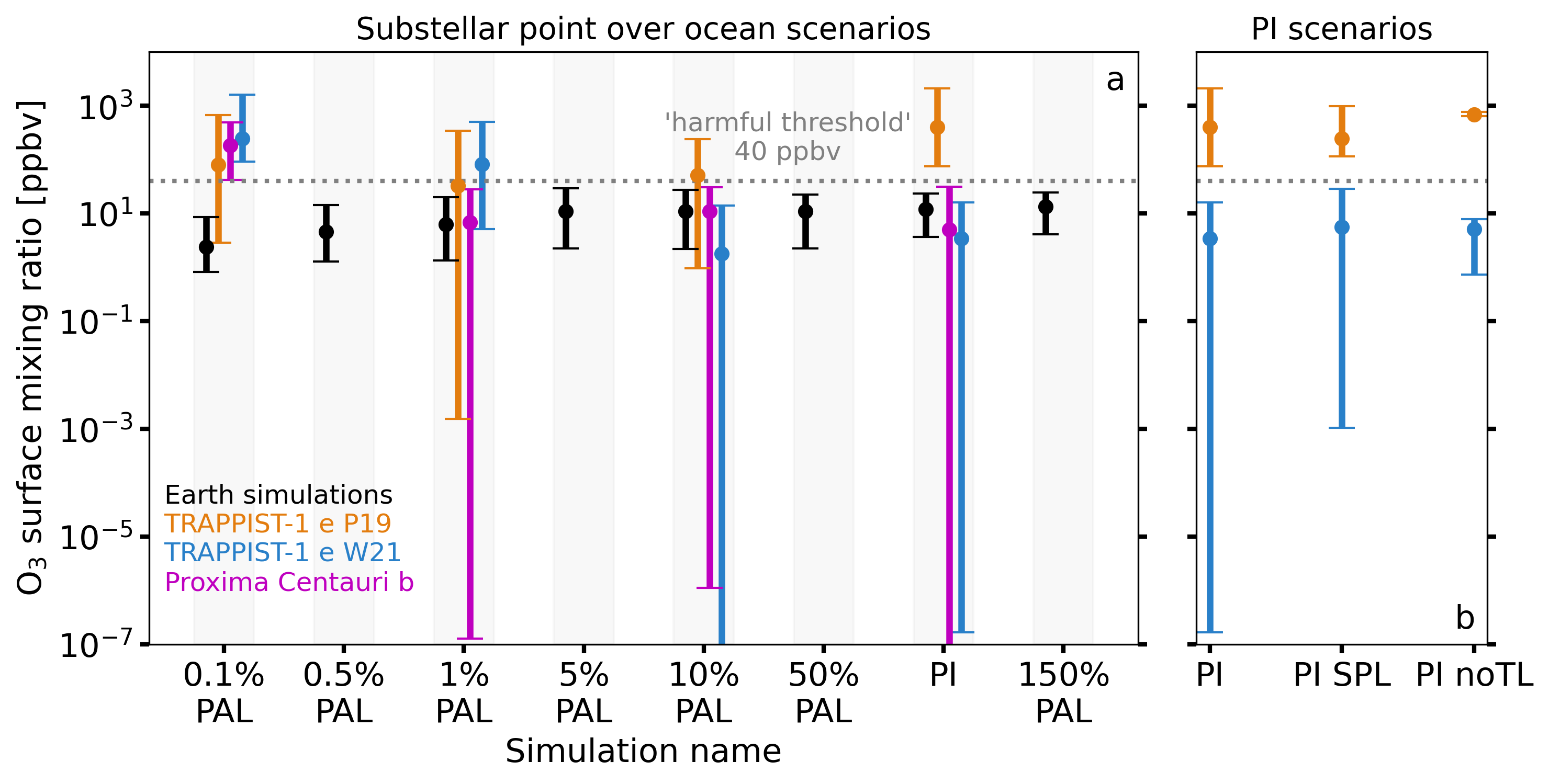}
    \caption{(\textbf{a}) Surface \ce{O3} mixing ratios are presented for the tidally locked P19 (orange) and W21 (blue) TP-1 e simulations, and the PCb simulations (magenta). All of these simulations have the substellar point placed over ocean. The dots show the mean mixing ratio, whilst the top bar shows the maximum, and the bottom bar shows the minimum surface \ce{O3} mixing ratios. Also shown in black are the time-averaged \ce{O3} mixing ratios of data taken from \cite{2022RSOS....911165C}. The horizontal axis indicates the simulations with a fixed \ce{O2} mixing ratio at the lower boundary, such that the horizontal ticks are categories rather than absolute values (the values are offset from each other for clarity). The grey dotted line indicates the 40 ppbv `harmful threshold', above which \ce{O3} surface mixing ratios are considered dangerous to some forms of life on Earth. (\textbf{b}) The pre-industrial scenarios are compared in the TP-1 e simulations. These include the substellar point over ocean (PI), over land (PI SPL), and the non-tidally locked cases (PI noTL).} 
    \label{Surface ozone comparison range figure}
\end{figure*}

\begin{figure*}[t!]
	\centering
	\includegraphics[width=0.9\textwidth]{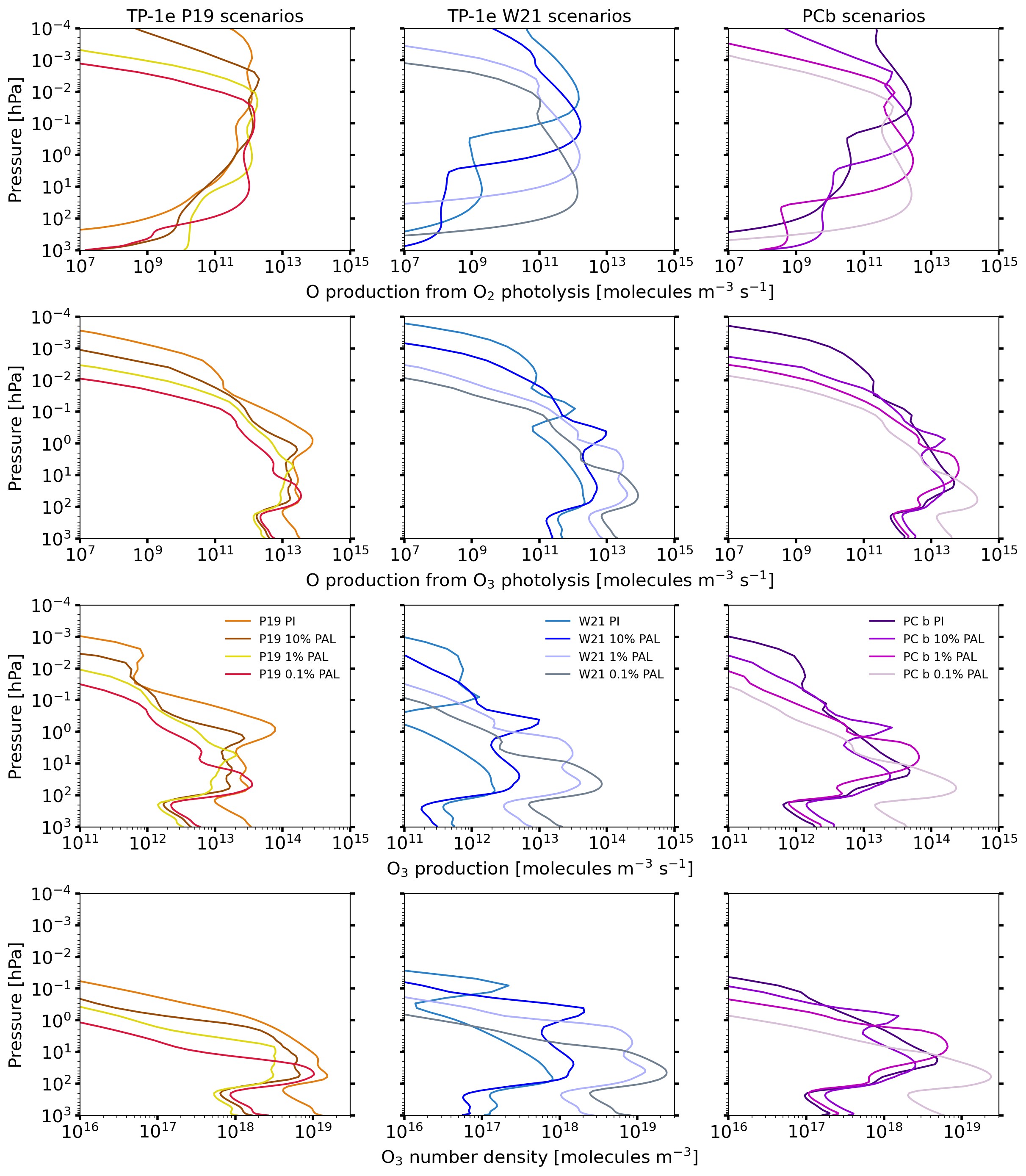}
    \caption{The production of O from \ce{O2} photolysis (top row) and \ce{O3} photolysis (second row) are shown against pressure, with \ce{O3} production (third row) and \ce{O3} number density also shown (bottom row). All profiles are time-averaged global means. The TP-1e PI, 10\% PAL, 1\% PAL, and 0.1\% PAL simulations are shown for both the TP-1 e P19 (orange, brown, yellow, and red, respectively) and TP-1 e W21 (light blue, blue, lilac, and grey, respectively) stellar spectra in the left and middle columns, respectively. The right column show the PCb PI (indigo), 10\% PAL (violet), 1\% PAL (magenta), and 0.1\% PAL (light pink) simulations.} 
    \label{Ox and O3 production and O3 density figure}
\end{figure*}

\begin{figure*}[t!]
	\centering
 \includegraphics[width=1\textwidth]{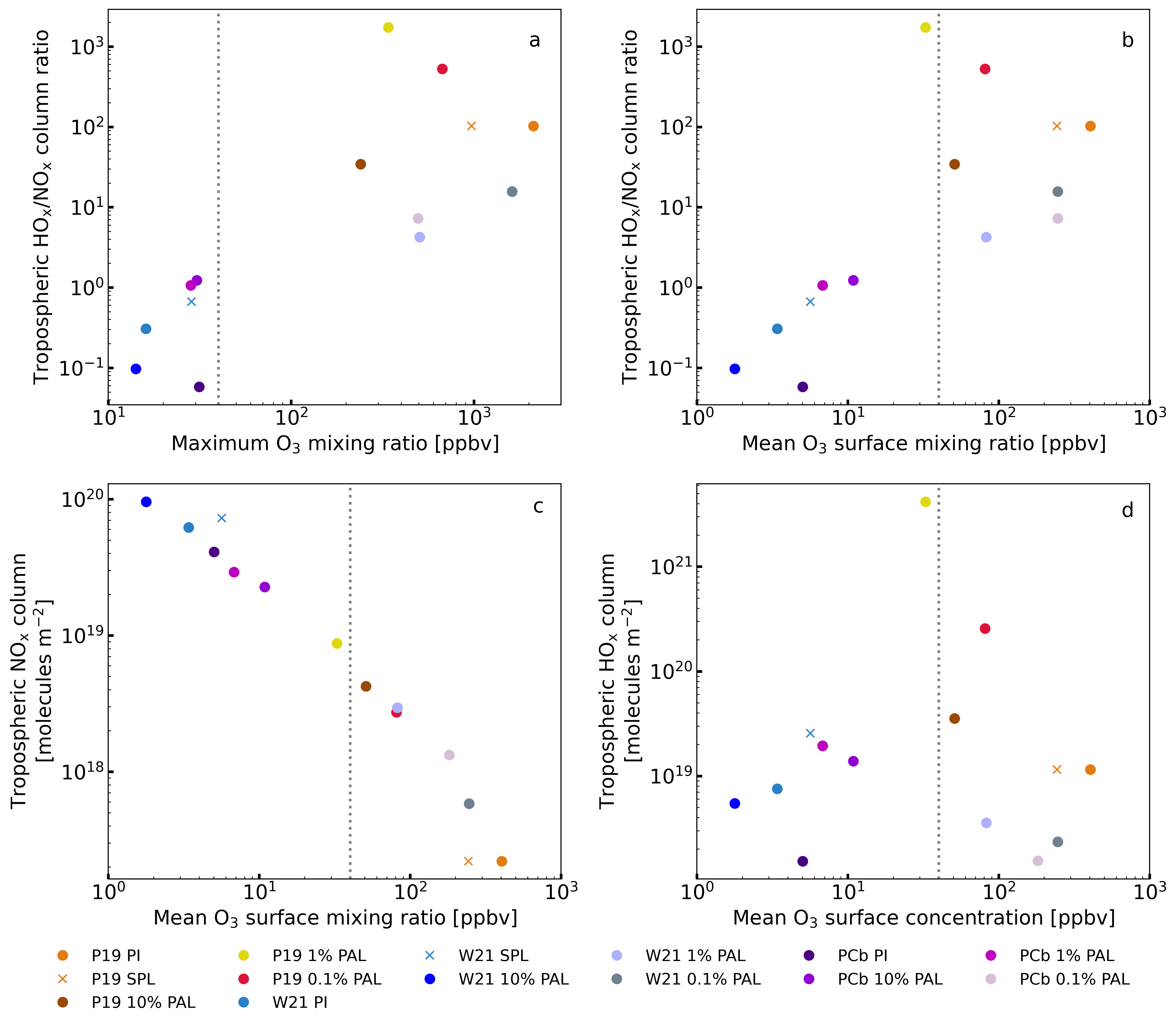}
    \caption{ The tropospheric HO\textsubscript{x}/NO\textsubscript{x} column ratio is shown against the maximum and mean surface \ce{O3} mixing ratio in each tidally locked simulation in panel \textbf{a} and \textbf{b}, respectively. The tropospheric NO\textsubscript{x} and HO\textsubscript{x} columns are shown against the mean \ce{O3} surface mixing ratio in panel \textbf{c} and \textbf{d}, respectively. The grey vertical dotted lines show the 40 ppbv cut off for harmful \ce{O3} levels. The tropospheric column is calculated as the column abundance of molecules between 120 hPa and the surface. Its units are molecules m\textsuperscript{-2}.} 
    \label{HOx NOx figure}
\end{figure*}

The bars in Fig.~\ref{Surface ozone comparison range figure}a indicate the full range of surface \ce{O3} mixing ratios in each of the tidally locked simulations where the substellar point is placed over ocean, with the global mean surface concentrations indicated by the points. Simulations of Earth (using WACCM6) at atmospheric \ce{O2} mixing ratios between 0.1\% -- 150\% PAL from data from \cite{2022RSOS....911165C} are given for comparison. Fig.~\ref{Surface ozone comparison range figure}b presents the same data for the TP-1 e PI scenarios. The TP-1 e W21 PI and 10\% PAL cases have a large range in surface \ce{O3} concentrations, spanning 7 and 8 orders of magnitude, respectively. All other TP-1 e simulations span 5 orders of magnitude or less, with the Earth simulations spanning only one order of magnitude. The PCb simulations span between 9 and 2 orders of magnitude. Whilst both the PI and PI SPL simulations with the P19 spectrum have harmful \ce{O3} mixing ratios at the surface, the mean surface \ce{O3} mixing ratio is reduced by 1.7 times when the substellar point is placed over land (PI SPL case). In contrast, the W21 PI SPL case has 2.0 times the mean surface \ce{O3} mixing ratio of the W21 PI case. Additionally, the noTL cases have a smaller range than both of the PI tidally locked cases, similar to the Earth simulations. These results imply that surface topography, the rotation rate, whether or not a diurnal cycle exists, and the position of the substeller point, will be important for modulating surface concentrations of biologically toxic gases such as \ce{O3}. For the PCb scenarios, the PCb 0.1\% PAL case reaches the largest \ce{O3} mixing ratio of 466 ppbv, and everywhere has mixing ratios exceeding 40 ppbv. None of the WACCM6 Earth simulations from \cite{2022RSOS....911165C} have time-averaged \ce{O3} mixing ratios at dangerous levels which is to be expected because industrial pollutants were not included in the simulations. If pollutants were included near urban areas, for instance, then harmful \ce{O3} levels would be localised, rather than on a planetary scale. The simulations which have surface \ce{O3} concentrations the most toxic to life are the PI, PI SPL, and PI noTL simulations with the P19 assumed spectrum. Quantitatively, this corresponds to the simulations where the global mean stratospheric \ce{O3} number density exceeds $7 \times 10\textsuperscript{18}$ molecules m\textsuperscript{-3}. All simulations with \ce{O2} concentrations at 0.1\% PAL have toxic mixing ratios exceeding 400 ppbv. 

\begin{figure*}[t!]
	\centering
	\includegraphics[width=1\textwidth]{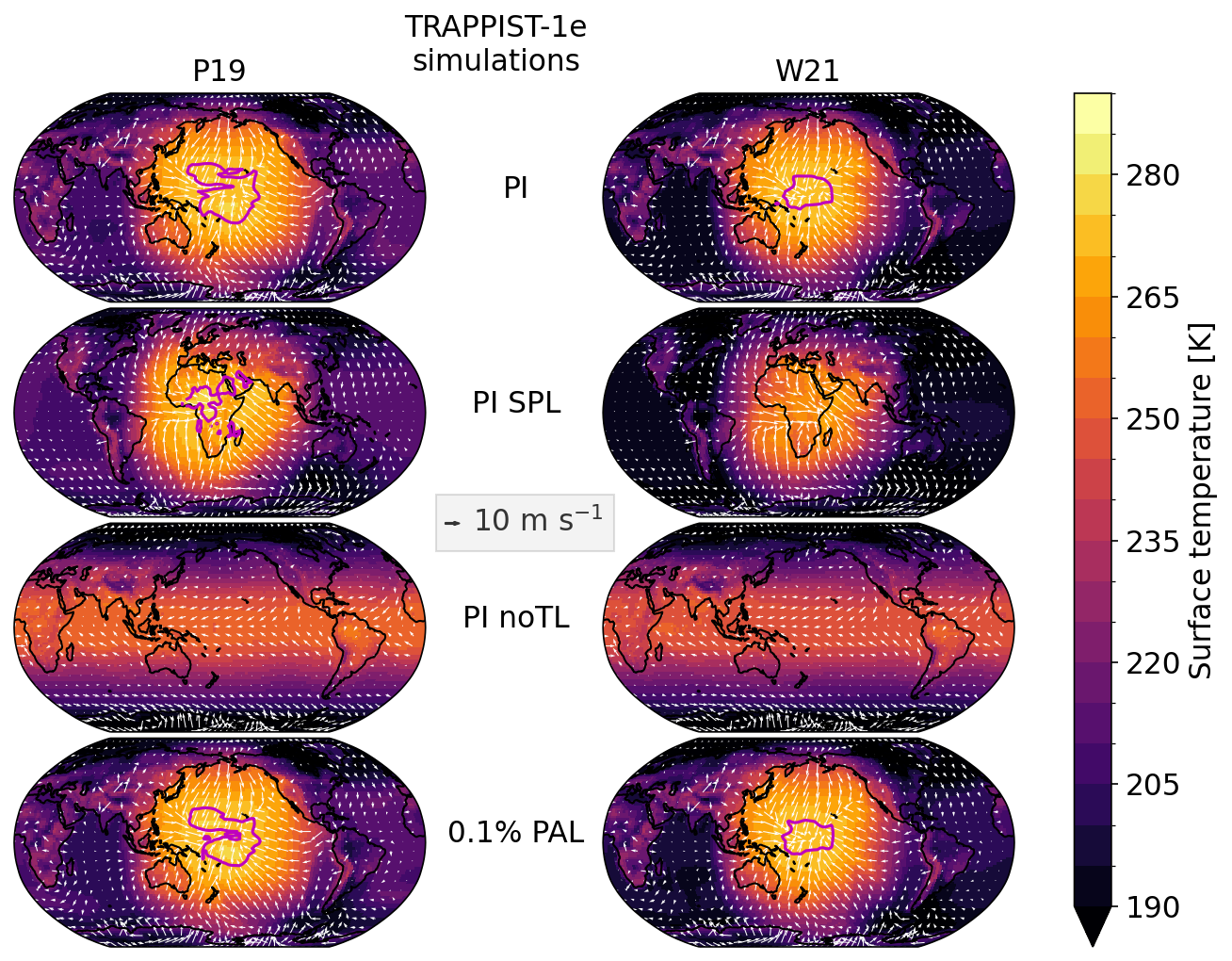}
    \caption{The surface temperature is displayed for the twelve TP-1 e simulations used in this work. The left column shows the P19 simulations and the right column shows the W21 simulations. From top to bottom: PI, PI SPL, PI noTL, and the 0.1\% PAL simulations are displayed. White arrows indicate the magnitude and direction of the surface winds. The magenta contours show surface temperatures of 273 K. The 10\% PAL and 1\% PAL simulations are not shown for brevity, but their surface temperatures are very similar to the PI and 0.1\% PAL cases. For scale, a 10 m s\textsuperscript{-1} arrow is shown in the middle of the figure.} 
    \label{Surface temperature figure}
\end{figure*}

\begin{figure*}[t!]
	\centering
	\includegraphics[width=1\textwidth]{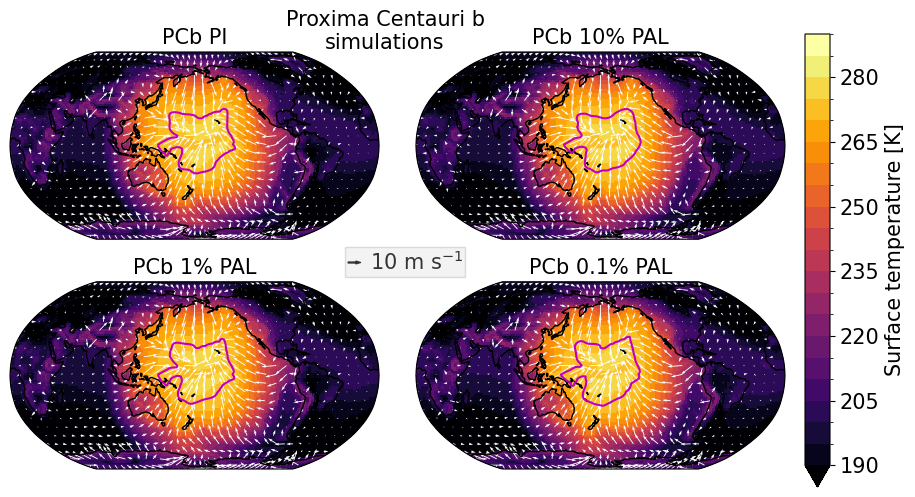}
    \caption{The surface temperature is displayed for the four PCb simulations used in this work: PCb PI (top left), PCb 10\% PAL (top right), PCb 1\% PAL (bottom left), PCb 0.1\% PAL (bottom right). White arrows indicate the magnitude and direction of the surface winds. The magenta contours show surface temperatures of 273 K. For scale, a 10 m s\textsuperscript{-1} arrow is shown in the middle of the figure.} 
    \label{Surface temperature figure PCb cases}
\end{figure*}

\begin{table}[b!]
\caption{The time-averaged and global mean dry deposition flux of \ce{O3} is given in terms of kg m\textsuperscript{-2} s\textsuperscript{-1}, for all the exoplanet simulations used in this work, as well as the Earth PI simulation. Dry deposition is a sink for atmospheric surface \ce{O3}. The time-averaged and global mean \ce{O3} surface mixing ratios are given. The fraction of the surface where \ce{O3} concentrations are under 40 ppbv and thus considered ``safe" is given for each simulated exoplanet scenario. The \ce{O3} concentrations vary every calendar month, so the fraction is given as a range over a 1-year period and as a percentage. Additionally, the global mean \ce{O3} column is given in Dobson Units (DU), where 1 DU is equal to $2.69 \times 10^{20}$ molecules m\textsuperscript{-2}.}
\centering
\begin{tabular}{@{}ccccc@{}}
\toprule
\multirow{2}{*}{Simulation} & \multirow{2}{*}{\shortstack{\ce{O3} dry deposition\\flux [kg m\textsuperscript{-2} s\textsuperscript{-1}]}} & \multirow{2}{*}{\shortstack{\ce{O3} surface\\ mixing ratio [ppbv]}} & \multirow{2}{*}{\shortstack{Fraction of surface with\\\ce{O3} mixing ratios $< 40$ ppbv}} & \multirow{2}{*}{\ce{O3} column [DU]} \\ \\
\hline
Earth PI & $2.1\times 10$\textsuperscript{-11} & 12 & N/A  & 297 \\
W21 PI & $1.0 \times 10$\textsuperscript{-12} & 3 & 100 -- 100\%  & 53 \\
W21 PI noTL & $4.3 \times 10$\textsuperscript{-15} & 5 & 100 -- 100\%  & 46 \\
W21 PI SPL & $1.3 \times 10$\textsuperscript{-14} & 6 & 100 -- 100\%   & 54 \\
W21 10\% PAL & $6.2 \times 10$\textsuperscript{-13} & 2 & 100 -- 100\%  & 154 \\
W21 1\% PAL & $1.8 \times 10$\textsuperscript{-11} & 82 & 1 -- 4\%  & 901 \\
W21 0.1\% PAL & $4.9 \times 10$\textsuperscript{-11} & 246 & 0 - 0\%  & 1227 \\
P19 PI & $9.1 \times 10$\textsuperscript{-11} & 404 & 0 -- 0\%  & 1289 \\
P19 PI noTL & $4.5 \times 10$\textsuperscript{-12} & 692 & 0 -- 0\%   & 1245 \\
P19 PI SPL & $4.3 \times 10$\textsuperscript{-11} & 243 & 0 -- 0\%  & 1098 \\
P19 10\% PAL & $1.2 \times 10$\textsuperscript{-11} & 51 & 12 -- 44\%  & 498 \\
P19 1\% PAL & $6.3 \times 10$\textsuperscript{-12} & 31 & 75 -- 83\%  & 260 \\
P19 0.1\% PAL & $1.5 \times 10$\textsuperscript{-11} & 81 & 4 -- 9\%  & 407 \\
PCb PI & $1.7 \times 10$\textsuperscript{-12} & 5 & 99 -- 100\%  & 179 \\
PCb 10\% PAL & $4.0 \times 10$\textsuperscript{-12} & 11 & 100 -- 100\%  & 134 \\
PCb 1\% PAL & $2.2 \times 10$\textsuperscript{-12} & 7 & 100 -- 100\%  & 256 \\
PCb 0.1\% PAL & $4.6 \times 10$\textsuperscript{-11} & 203 & 0 -- 1\%  & 790 \\
\hline
\end{tabular}
\label{Dry dep table}
\end{table}

\subsection{Cause of ozone production at the surface}

It is important to state here that initially in the simulations, \ce{O3} was present in Earth-like quantities throughout the troposphere, and no toxic concentrations existed at the start of the tidally locked simulations. The \ce{O3} profile depends on UV radiation, \ce{O2} number density, \ce{O3} production rates, \ce{O3} loss, and the transport of \ce{O3}. Fig.~\ref{Ox and O3 production and O3 density figure} shows the photolysis rates of \ce{O2} leading to O production (reactions \ref{O2 photolysis reaction 1} and \ref{O2 photolysis reaction 2}), the photolysis rates of \ce{O3} leading to O production (reactions \ref{ozone photolysis 1} and \ref{ozone photolysis 2}), the production rates of \ce{O3} (reaction \ref{Ozone production reaction}), and the \ce{O3} number density in each simulation. These quantities are important for understanding where \ce{O3} is produced, and its resulting number density. \ce{O2} has an approximately constant mixing ratio up until the homopause where gases start to diffusively separate, but in contrast, the mixing ratio of O increases with altitude until the homopause.

In the W21 simulations, as \ce{O2} decreases, the total amount of \ce{O3} in the atmosphere, and at the surface, increases. The opposite is generally true in the P19 simulations, although there is an increase in surface \ce{O3} between 1\% PAL and 0.1\% PAL. This difference between the P19 and W21 scenarios occurs due to the weak UV radiation in the W21 simulations, and the pressure dependency on the reaction which produces \ce{O3} (reaction \ref{Ozone production reaction}). When the peak of \ce{O2} photolysis occurs at higher altitudes and thus lower pressure, then O does not react with \ce{O2} as quickly to produce \ce{O3} compared to the rate lower in the atmosphere where the density of the third body, M, is higher. In the P19 cases, whilst the UV can penetrate deeper into the atmosphere when the concentration of \ce{O2} is reduced in the simulations, the availability of \ce{O2} becomes the limiting factor for the production of \ce{O3}, instead of UV radiation. In the PCb cases, there is an intermediate amount of UV radiation compared to the P19 and W21 TP-1 e cases. These PCb cases follow the same trend to the TP-1 e W21 scenarios, with the surface \ce{O3} increasing as \ce{O2} is reduced. 

The destruction of \ce{O3} plays an important role in these atmospheres too. Photolysis of \ce{O3} is not counted as a loss of \ce{O3}, because the O produced quickly cycles back to produce \ce{O3}. The peak in \ce{O2} photolysis, \ce{O3} production, and \ce{O3} number density, occurs in each simulation at pressures less than 100 hPa (above the troposphere). \ce{O3} formation also takes place in the troposphere because \ce{O3} photolysis there is fast due to \ce{O3} being present in relatively high quantities. \ce{O3} is being destroyed (by \ce{HO}\textsubscript{x} and \ce{NO}\textsubscript{x} catalytic cycles) and remade in the troposphere, but compared to above the troposphere, its production via \ce{O2}, \ce{CO2}, \ce{NO2}, or \ce{H2O} photolysis is significantly slower. In the simulations that have toxic quantities of \ce{O3}, tropospheric destruction of \ce{O3} is dominated by \ce{HO}\textsubscript{x} catalytic cycles rather than \ce{NO}\textsubscript{x} catalytic cycles. When \ce{O3} is below harmful levels, \ce{NO}\textsubscript{x} catalytic cycles (predominantly \ce{NO} and \ce{NO2}) dominate over \ce{HO}\textsubscript{x} catalytic cycle destruction of \ce{O3}. In other words, when \ce{NO} and \ce{NO2} are significantly depleted in the troposphere, \ce{O3} is able to accumulate to harmful levels. The tropospheric column amount of \ce{NO}\textsubscript{x} exhibits negative correlation with the mean surface \ce{O3} mixing ratio, as shown in Fig.~\ref{HOx NOx figure}. Therefore, the smog mechanism is not the reason for harmful levels of \ce{O3}, in contrast to modern day Earth.

The atmospheric temperatures and dynamics influence the abundance and distribution of \ce{O3}. Fig.~\ref{Surface temperature figure} shows the surface temperatures (colour map) and surface winds (arrows) in the TP-1 e models, with Fig.~\ref{Surface temperature figure PCb cases} showing the same in the PCb models. Several TP-1 e tidally locked simulations have substellar points with temperatures above 273 K, and surface winds converging towards this point (associated with the day-side upwelling at the sub-stellar point), with some of the lowest surface mixing ratios also found near the substellar point. Away from the substellar point, temperatures drop below freezing and can be as low as 170 K on the nightside. The PCb cases have similar surface temperatures and winds because the total irradiance is quantitatively similar to the TP-1 e scenarios (0.65 $S_0$ versus 0.66 $S_0$). The cold temperatures result in reduced destruction from catalytic cycles (see reaction \ref{Catalytic cycles}) which proceed slower at lower temperatures (e.g., from HOx and NOx families), allowing \ce{O3} to persist in relatively high quantities. 

\begin{figure*}[t!]
	\centering
	\includegraphics[width=1\textwidth]{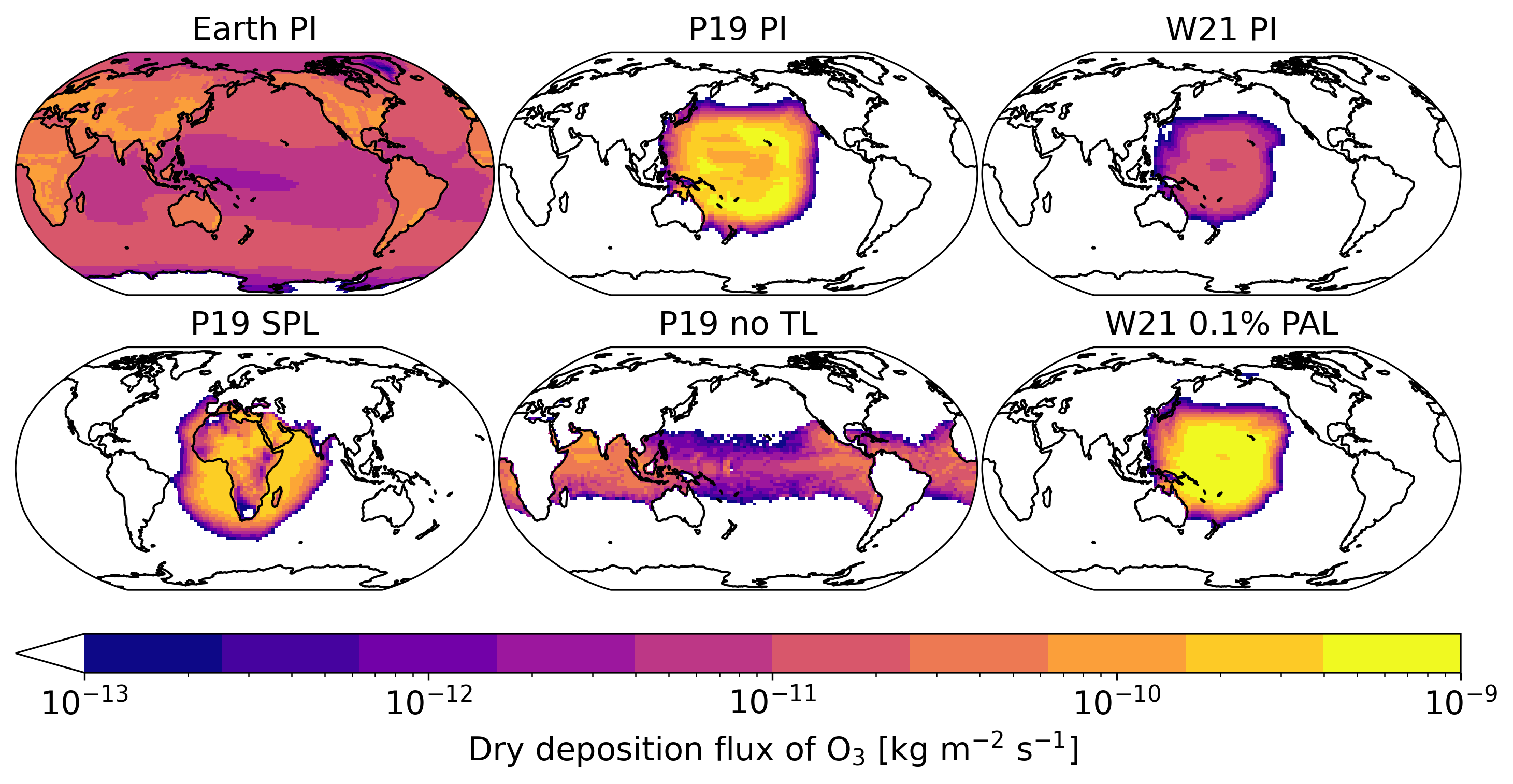}
    \caption{The dry deposition flux of \ce{O3}, given in terms of kg m\textsuperscript{-2} s \textsuperscript{-1}, is plotted for six simulations: The Earth PI simulation, the TRAPPIST-1 e P19 PI, SPL, noTL, and W21 PI and 0.1\% PAL simulations. Yellow indicates relatively large amounts of dry deposition, whilst red indicates relatively low amounts. The white areas indicate regions where the dry deposition flux of \ce{O3} is below $10^{-13}$ kg m\textsuperscript{-2} s \textsuperscript{-1}.} 
    \label{Dry deposition figure}
\end{figure*}

Fig.~\ref{Dry deposition figure} shows the dry deposition flux of \ce{O3} for some of the TRAPPIST-1 e simulations and the Earth PI simulation, and Table~\ref{Dry dep table} shows the global mean dry deposition flux for the Earth PI, TRAPPIST-1 e, and Proxima Centauri b simulations. Dry deposition over snow and ice is slow compared to that over other surfaces \citep{2000AtmEn..34.2261W,2007ACP.....7...15H,barten2023low}. Additionally, marine surface deposition is slower than land deposition when plant stomata are available to take up \ce{O3} \citep{ainsworth2012effects}. If surface plants do not exist or have died from prolonged \ce{O3} exposure \citep{rich1964ozone, bytnerowicz1993detecting, sandermann1996ozone, rao2001physiology, 2023ER....236k6816R}, then dry deposition would occur more slowly, which results in even more \ce{O3} build-up. Even if there are no surface or gaseous molecules for \ce{O3} to interact with, the surface thermal decomposition of \ce{O3} can take place and its importance may vary depending on the surface type, although this process is currently not sufficiently understood \citep{2009AtmEn..43.5193F, 2020RvGeo..5800670C}. The Earth PI simulation has a global mean dry deposition flux of $1.6\times 10$\textsuperscript{-11} kg m\textsuperscript{-2} s\textsuperscript{-1}. All TRAPPIST-1 e and Proxima Centauri b simulations have reduced loss rates when compared to this, apart from the P19 PI, W21 1\% PAL, W21 0.1\% PAL and PCb 0.1\% PAL simulations. Despite these cases having relatively high rates of dry deposition, all retain harmful concentrations of surface \ce{O3}. For the Earth PI case, most \ce{O3} is deposited over land (see Fig.~\ref{Dry deposition figure}). For the exoplanet simulations, the majority of \ce{O3} is deposited near the substellar point, regardless of whether it is placed over land or ocean. The dry deposition flux around the substellar point contributes to the relatively reduced \ce{O3} concentrations at the substellar point (e.g., see the P19 10\% and 0.1\% PAL simulations and W21 1\% PAL and 0.1\% PAL simulations in Fig.~\ref{Surface ozone figure - TP-1 e}). The noTL cases have globally-averaged dry deposition rates that are slower than their PI tidally locked counterparts by a factor of 236 and 20 for the W21 and P19 scenarios, respectively. 

To summarise the \ce{O3} distribution in the simulations, \ce{O3} is made primarily in the middle atmosphere on the modelled dayside and the \ce{O3} column maximises at the poles \citep{cooke2023degenerate}. \ce{O3} loss on the nightside and at the poles is slower than the dayside due to relatively low temperatures and a lack of photolysis producing the molecules that become involved in \ce{O3} destroying catalytic cycles. \ce{O3} is lost at the surface due to dry deposition, but the flux is not large enough to mitigate for the dangerous concentrations of \ce{O3}. As an example, in the P19 PI case, the global mean dry deposition flux increases by $\approx 4.3$ times greater than the Earth PI case, with a corresponding increase in surface \ce{O3} concentrations of $\approx 33.7$.  In the tidally locked cases, the \ce{O3} chemical loss rate at the surface is approximately 2 -- 29 times less than the peak stratospheric loss rate. Surface winds (of order 10 m s\textsuperscript{-1}), which are much stronger than vertical winds (of order 0.1 m s\textsuperscript{-1}), transport \ce{O3} across the surface. 

We hypothesise that \ce{O3} is transported from where it is produced in the dayside stratosphere to the nightside and towards the poles as discussed in \cite{2023MNRAS.526..263B}, who simulated Proxima Centauri b assuming an initial condition of a modern-Earth atmosphere. The difference between our work and \cite{2023MNRAS.526..263B} is, in several of our simulations, \ce{O3} accumulates to harmful and lethal quantities. With loss processes in the troposphere less effective than in the stratosphere, the lifetime of surface \ce{O3} increases and enables a build-up of toxic \ce{O3} levels. The atmospheric transport throughout the entire atmosphere will be explored in a follow-up paper to confirm whether such a scenario is occurring in the simulations.

\section{Discussion}
\label{Discussion}

\subsection{Prior results for toxic levels of ozone}

As far as can be discerned from the presented data (e.g. globally averaged vertical profiles), other 3D simulations of oxygenated exoplanets \citep{2016EP&S...68...96P, 2017ApJS..231...12W, 2019ApJ...886...16C, 2020MNRAS.492.1691Y}, excluding \cite{2022MNRAS.517.2383B}, have not produced surface \ce{O3} mixing ratios above 40 ppbv. This is likely due to the investigated scenarios which differ between each work, although model differences will be important \citep{2023RSOS...1030056J, ji2024correlated}. Figure 5 in \cite{2022MNRAS.517.2383B} shows dayside mixing ratios of \ce{O3} reaching $\approx 45$ ppbv at the surface. It is worth noting here that all of these studies, including our simulations, assume a surface pressure of 1 bar (1,000 hPa), so that a mixing ratio of 40 ppbv at 1,000 hPa surface pressure corresponds to a number density of $1.0\times10^{18}$ molecules m\textsuperscript{-3} at 288 K. \cite{2021NatAs...5..298C} used WACCM4 and \cite{2023MNRAS.518.2472R} used the UM (two 3D chemistry-climate models) to investigate the impact of flares and coronal mass ejections on terrestrial exoplanets. The flares significantly perturbed \ce{O3} concentrations\footnote{The amount of atmospheric \ce{O3} decreased by a factor of 3 in \cite{2021NatAs...5..298C} for active M dwarf stars, whereas total atmospheric \ce{O3} increased by up to a factor of 20 in \cite{2023MNRAS.518.2472R}.}, but the changes were in the middle and upper atmosphere and the flares did not cause surface concentrations to exceed 40 ppbv. However, there could be specific cases where flares act to increase \ce{O3} surface concentrations to harmful levels, likely depending on the atmospheric properties, incoming flare strength, and flare frequency.

1D photochemical models have also simulated the impact of flares, coronal mass ejections, and cosmic rays on atmospheric chemistry. The calculations in \cite{2010AsBio..10..751S}, \cite{2012AsBio..12.1109G}, \cite{2016A&A...585A..96T}, and \cite{2019AsBio..19...64T}, did not demonstrate surface \ce{O3} mixing ratios reaching biologically toxic quantities. The same is true for abiotic \ce{O2} production simulations \citep{2007A&A...472..665S, 2015ApJ...812..137H, 2018ApJ...866...56H}, although it is unclear in the 100 bar atmosphere from \cite{2016ApJ...819L..13S} because surface \ce{O3} mixing ratios are not shown (their figure 1), but \ce{O3} is at 1 ppbv by 10 km and its mixing ratio is strongly decreasing with decreasing height. Other 1D photochemical modelling results for habitable zone exoplanets have shown that \ce{O3} mixing ratios may exceed harmful levels \citep{2018ApJ...862...69K, 2019ApJ...887..194F, 2020ApJ...904...10K, 2021ApJ...908..164K}. For example, \cite{2018ApJ...862...69K} used a 1D photochemical model \citep[EXO-Prime;][]{2010ApJ...708.1162K} to simulate Earth-like exoplanets with various surface pressures of 0.3, 1.0, 1.5, and 2.0 bar, orbiting white dwarf stars (stellar effective temperatures of 4000, 5000, and 6000 K). Almost all simulations had surface \ce{O3} below the 40 ppbv threshold, with only a single simulation (0.3 bar surface pressure and 4000 K stellar effective temperature) exceeding it. However, if it is the concentration (number of molecules per unit volume), rather than the mixing ratio (fractional concentration), which is important, one must take into account the surface density of the atmosphere. Using this criterion, a few more simulations from \cite{2018ApJ...862...69K} would be close to $1.0\times10^{18}$ \ce{O3} molecules m\textsuperscript{-3} and therefore potentially dangerous to any surface life present. This was the case for simulated atmospheres around white dwarf stars, however, the 1 bar atmospheres in \cite{2020AJ....160..225K} which were simulated around red giant stars did not surpass dangerous surface \ce{O3} concentrations. Regarding smog (see Section \ref{Introduction section} and Reactions \ref{smog reaction 1} -- \ref{smog reaction 3}), \cite{2021ApJ...908..164K} used the 1D photochemical model which is part of `Atmos' \citep[see][]{2016AsBio..16..873A, 2019ApJ...873L...7A} to simulate varying surface-to-atmosphere fluxes of \ce{NO2} to test whether it could be used as a signature that extraterrestrial technology existed on an exoplanet. In two simulations (using a Sun-like star and a K6V star with $20\times$ the present Earth flux of \ce{NO2}), the surface \ce{O3} mixing ratios were $\approx 90$ and $\approx 100$ ppbv, respectively. Alongside modern day pollution on Earth due to NOx emissions from vehicles, this study by \cite{2021ApJ...908..164K} demonstrates that the surface fluxes of molecules will be important for determining whether ground level \ce{O3} concentrations reach concerning levels for life. \cite{2019ApJ...887..194F} simulated the habitable zone exoplanets TRAPPIST-1 e, f, and g, using the 3D model LMD-G and then performed terminator photochemistry simulations with Atmos. The surface \ce{O3} concentrations are $< 40$ ppbv for TRAPPIST-1 e, but are $\sim 60$ ppbv and $\sim 120$ ppbv for TRAPPIST-1 f and g, respectively. In their simulations, surface \ce{O3} increases with decreasing illumination. The simulations presented here predict, for the P19 PI case, the highest ozone surface concentrations ($7.8\times10^{19}$ molecules m\textsuperscript{-3}) compared with other results for exoplanets simulated in the literature. 

WACCM6 predicts lower concentrations of \ce{O3} for various oxygenation states (0.1 -- 150\% PAL) for Earth when compared to 1D models \citep{2022RSOS....911165C, 2023RSOS...1030056J}, so there is the question of why it now predicts higher concentrations of \ce{O3} when compared to 1D models for M dwarf stars. There will likely be many reasons the results differ \citep[e.g., see][]{2023RSOS...1030056J}, but we suspect that the main causes are differences in atmospheric transport and temperatures. The atmospheric temperatures in \cite{2022A&A...665A.156K} with the M5V star are warmer below 30 km when compared to the WACCM6 simulations. The same is true in the M3V and M8V simulations of \cite{2018ApJ...854...19R}, the GJ 436 and GJ 876 simulations in \cite{2015ApJ...809...57R}, the \cite{2018AsBio..18..856G} simulations around AD Leo (M3.5V), the TRAPPIST-1 e simulations in \cite{2021MNRAS.505.3562L}, the Proxima Centauri b simulations in \cite{2020ApJ...893...12S}, and the 1 bar habitable Proxima Centauri b simulations in \cite{2018AsBio..18..133M}. Note that the desiccated Proxima Centauri b atmospheres in \cite{2018AsBio..18..133M} do have harmful levels of \ce{O3}, but these would not be considered habitable due to the lack of \ce{H2O}. 

As an example of the temperature differences, global mean surface temperatures are approximately 220 -- 230 K in the WACCM6 tidally locked simulations, compared to a surface temperature of $\approx 310$  K in \cite{2022A&A...665A.156K}, 260 -- 280 K in \cite{2020ApJ...893...12S}, and 273 K in \cite{2018AsBio..18..133M}. This means that below 30 km, where most of the atmospheric \ce{O3} resides, loss rates will be slower and formation will be faster when compared to the 1D models. Furthermore, in 1D models, there is constant illumination, but in 3D models, \ce{O3} is transported to the nightside (where there is no direct starlight) and down to the surface \citep{2023MNRAS.526..263B}. The TRAPPIST-1 e 1D photochemical simulations in \cite{2020ApJ...898L..33P} used temperature and pressure profiles from 3D simulations of TRAPPIST-1 e \citep{2019ApJ...887..194F}. Here, the deviations from our results may be due to the synthetic BT-Settl spectrum \citep[a model of stellar atmospheres;][]{2013A&A...556A..15R} used by \cite{2020ApJ...898L..33P}, and the fact that photochemistry was performed at the terminator. Between all of these simulations, there will be differences in photochemical cross sections and chemical schemes, as well as the UV spectra used and the total instellation, but temperatures and transport may largely explain the substantial discrepancies both in predicted surface \ce{O3}, and \ce{O3} columns, when compared to similar 1D model simulations.

With regards to why WACCM6 predicts higher surface concentrations of \ce{O3} compared to the Unified Model will require a detailed intercomparison. For now, we speculate at least one difference is the use of a slab ocean with no ice formation in the Unified Model, such that \ce{O3} will be lost in greater numbers on the nightside than would occur in reality. When simulating Proxima Centauri b with a slab ocean, \cite{2020MNRAS.492.1691Y} and \cite{2022MNRAS.517.2383B} assumed a dry deposition velocity of 0.05 cm s\textsuperscript{-1} based on previous work \citep{1995JGR...10020999G, 1999JGR...10423761G}. The global mean dry deposition velocities for \ce{O3} in our tidally locked ocean substellar point cases are approximately 4 times slower, and slower still when considering the ``noTL" and ``SPL" cases. \ce{O3} dry deposition depends on multiple interlinked parameters which are poorly known \citep{2017Atmos...8..175E}, so whether existing dry deposition parameterisations can be used for exoplanet simulations is not well known.

\subsection{Modelling limitations}

The model used for the atmospheric simulations is an important factor in these predictions because varying parameterisations and chemical schemes will impact the results. WACCM6 is a model that is tuned to Earth's atmosphere, land, ice and ocean. WACCM6 accounts for scattering longward of 200 nm, but it doesn't account for scattering in the Schumann-Runge bands (175 -- 192 nm; these wavelengths photolyse \ce{O2} above $\sim 80$ km in Earth's atmosphere), becoming pertinent for Earth-like simulations at \ce{O2} mixing ratios of 1\% PAL or less \citep{2023RSOS...1030056J}. In our simulations, WACCM6 accounts for absorption in the Schumann-Runge bands from \ce{O3}, \ce{O2}, \ce{CO2}, and \ce{H2O}. The integrated flux in the Schumann-Runge Bands is 1.15, 656, and 26.6 times lower than Earth, for the TP-1 e P19, TP-1 e W21, and PCb cases, respectively. Even with the effects of scattering included, it seems still possible that harmful \ce{O3} surface concentrations could form on the surface of terrestrial exoplanets because all of the P19 simulations have dangerous concentrations of surface \ce{O3}. Moreover, the W21 simulations have a relatively low amount of radiation in the Schumann-Runge bands, yet toxic \ce{O3} concentrations are found in the W21 1\%, and 0.1\% PAL cases. We tested these assumptions in the 0.1\% PAL cases by reducing the incoming photon flux in the W21 and P19 spectra by a factor of $10^5$. The total integrated \ce{O2} photolysis in the P19 0.1\% PAL case reduced by a factor of 1.6, and in the W21 case the decrease was negligible. The decrease in \ce{O2} photolysis in the P19 0.1\% PAL case primarily takes place between 10 and 0.01 hPa. As seen from the results presented in figure 7 in \cite{2023RSOS...1030056J}, scattering becomes important in the 0.1 -- 10\% PAL simulations at altitudes below 10 hPa. Therefore we expect our conclusions to be unaffected by including scattering in the Schumann-Runge band parameterisation in WACCM6; however, the test should still be done when that parameterisation is eventually updated.
 
\subsection{Atmospheric evolution}

We have assumed planetary conditions based on the last 2.4 billion years of Earth's history, where the atmospheric pressure has been $\sim 1$ bar and atmospheric \ce{O2} has varied between $1000\times$ less than the present atmospheric level to $\sim 1.5\times$ greater \citep{2019MinDe..54..485L,2020SciA....6.1420C, 2020PreR..343j5722S, 2021AsBio..21..906L}. However, for a habitable zone tidally locked exoplanet around an M dwarf star, is such a scenario realistic?

Firstly, atmospheres with high \ce{O2}, \ce{CO2}, or \ce{H2O} mixing ratios and sufficient UV may develop an \ce{O3} layer close to the surface because photodissociation could produce \ce{O2} and O molecules which lead to \ce{O3} formation. In previous work simulating a 1 bar atmosphere around the host star GJ 876, \ce{O2} amounts comparable to the Proterozoic Earth were produced in abiotic scenarios \citep{2014ApJ...792...90D, 2014E&PSL.385...22T}. When \ce{CO2} is photolysed it produces \ce{CO} and \ce{O}, but lightning flashes can produce NO, which can catalyse the recombination of \ce{CO} and \ce{O} (instead of O going on to produce \ce{O2}), meaning that abiotic \ce{O2} may not increase to detectable levels \citep{2018ApJ...866...56H}. However, on some exoplanets lightning may not be sufficient to prevent \ce{O2} build-up \citep{2024arXiv240213682B}.

Oxygenic photosynthesis on Earth, utilising chlorophyll as a pigment, requires photosynthetically active radiation (PAR) between $400–700$ nm \citep{1996AgFM...78..121A}. Due to the luminosity of M dwarf stars peaking at longer wavelengths when compared to the Sun, the availability of PAR may be lacking on habitable M dwarf exoplanets. Nevertheless, oxygenic photosynthesis is thought to be possible on exoplanets orbiting M dwarf stars \citep{2017IJAsB..16....1G, 2020Life...11...10C, 2023MNRAS.526.2265D}, although some exoplanets may be limited by a lack of photons reaching the surface \citep{2018ApJ...859..171L} and this could mean that anoxygenic photosynthesis (where oxygen is not a product of photosynthesis) is preferentially selected \citep{2023MNRAS.526.2265D}.  In particular, the results from \cite{2019MNRAS.485.5924L} showed how planets orbiting stars with masses below $0.13\ M_\odot$ may have difficulties in producing \ce{O2} through oxygenic photosynthesis, with Proxima Centauri and TRAPPIST-1 falling under this limit. However, the results of \cite{2021MNRAS.505.3329C} contrasted to \cite{2019MNRAS.485.5924L}, and whilst TRAPPIST-1 f and g might be energy limited \citep{2018ApJ...859..171L, 2021MNRAS.505.3329C}, TRAPPIST-1 e may be able to host an Earth-like biosphere. Furthermore, cyanobacteria can use radiation longward of 700 nm to photosynthesize \citep{gan2015adaptive, 2020Life...11...10C}, and a recent study has also shown this in more complex vegetation \citep{zhen2022photosynthesis}. A new metric quantifying the photo-absorption rate of photosynthetic pigments around different stars, suggests M8V stars could activate photosynthesis \citep{2022Univ....8..624M}. Moreover, cyanobacteria can survive and photosynthesize in caves and under low light conditions \citep{1997MEPS..149..255H, behrendt2020life, 2023FrASS..1007371J}. Regardless of the possibility of photosynthesis on M dwarfs, \ce{O2} production needs to surpass \ce{O2} sinks in order to produce an oxygenated atmosphere \citep{2018ApJ...859..171L}.

It is hypothesized that the oxidation of Earth's lithosphere (the solid outer layer of a planet) occurred due to two primary factors: oxygenic photosynthesis from cyanobacteria over millions of years, and hydrogen escape to space resulting from water photolysis \citep{catling2001biogenic, 2013ChGeo.362...26Z}. Modelling an Earth-like planet orbiting the star AD-Leo (M3.5V), \cite{2018AsBio..18..856G} found that it may be possible for a Great Oxidation Event \citep[see e.g.,][]{2014Natur.506..307L, gumsley2017timing, poulton2021200} to occur earlier in the history of the modelled planet when compared to Earth. If \ce{O2} increases and results in large amounts of \ce{O3} at the surface, because \ce{O3} is a strong oxidiser, it could speed up the oxidation of Earth's lithosphere, reducing the sinks of \ce{O2} and further enabling the generation of an oxygenated atmosphere.

\subsection{Habitability feedback}

Continuing with this scenario in mind, there is the potential for biological feedback. Surface \ce{O3} may build-up to toxic concentrations and kill organisms that produce \ce{O2} on the surface, subsequently reducing the production of \ce{O2} until \ce{O3} reaches safe levels again. Even so, we argue here that \ce{O3} could exceed lethal surface concentrations and a photosynthetic biosphere could still be safely present in the ocean and under ice.

Firstly, refuge from the dangerous \ce{O3} concentrations may be found in any liquid water ocean present, because \ce{O3} has low solubility in water \citep{egorova2015ozone}. Although it is possible for \ce{O3} to form in water which is irradiated by the Sun \citep{lushchak2011environmentally}, \ce{O3} then quickly decomposes. Additionally, for disinfection applications, \ce{O3} has to be artificially inserted into water \citep[e.g., via bubble diffusion;][]{wert2017effect} such that it seems unlikely that marine life would be adversely affected by high surface concentrations of \ce{O3} in the air. The known effects of \ce{O3} exposure on life are limited to observations of terrestrial organisms, and extraterrestrial organisms may adapt to survive in an atmosphere with high surface \ce{O3} mixing ratios.

On Earth, phytoplankton blooms have been found under Arctic \citep{1997JMS....11..111S, arrigo2012massive, 2020JGRC..12516211C} and Antarctic sea ice \citep{mcminn2007spring}, algae are found at varying depths \citep{norris1991deep, pritchard2013survival, borlongan2017photosynthetic}, and lichens \citep{kappen1993plant}, aquatic plants \citep{adams1974depth, 2007ECSS...73..551C}, and seaweed \citep{ramus1976correlation, huovinen2013photosynthetic}, can photosynthesize below the surface. The long day length during the summers at each pole on Earth can increase the rate of primary productivity \citep{henshaw2002annual}. Given phytoplankton's contribution to roughly half of Earth's primary productivity \citep{1998Sci...281..237F} and the potential for subsurface photosynthesis on tidally locked exoplanets with constant day side illumination, significant oxygen accumulation may occur without surface life. The amount of ice coverage and available area of open ocean may limit the \ce{O2} flux to the atmosphere, although gaseous diffusion through sea ice is possible \citep{2007LimOc..52.1367D,loose2011sea, 2011TellB..63...23L, 2012IzAOP..48..538B}, just slower when compared to open water \citep{2012IzAOP..48..538B}.

In summary, whether due to biological or abiotic production, \ce{O2} could be produced in quantities large enough to induce toxic concentrations of \ce{O3} at the surface. If life is present on such a planet, there is then the potential for biogeochemical feedback as organisms are hindered or destroyed by \ce{O3}. We encourage simulations of these scenarios using biogeochemical models to quantitatively determine the possible outcomes we have discussed.

\subsection{Future work}

Future work should aim to determine the parameter space (in UV irradiation, composition, and atmospheric pressure) for which detrimental levels of surface level \ce{O3} may occur. Ideally, 1D photochemical models would be used for this as they are less computationally expensive than 3D models, and the use of 1D photochemical models is indeed a viable investigation for non-tidally locked exoplanets. However, for tidally locked exoplanets, where \ce{O3} is transported to the nightside and shielded from destruction \citep{2016EP&S...68...96P, 2020MNRAS.492.1691Y, 2023MNRAS.526..263B}, 3D chemistry-climate models will be needed to predict where surface \ce{O3} concentrations maximise due to atmospheric dynamics, and climatology from varied ocean and land configurations \citep{2021ApJ...910L...8Z, 2022MNRAS.513.2761M}. The ocean salinity could be varied to determine the affect on the sea ice distribution \citep{2022GeoRL..4995748O}, which is important for habitability estimates and \ce{O3} dry deposition calculations. Simulations where only the ocean model is changed between dynamic ocean, slab ocean with ice formation \citep[e.g.,][]{2022GeoRL..4995748O}, and slab ocean without ice formation \citep[e.g.,][]{2022MNRAS.517.2383B}, would be useful to determine how important these factors are when considering toxic \ce{O3} at the surface. Additionally, one could explore various topographies, gravitational accelerations, and several different atmospheric compositions (e.g., \ce{H2O} or \ce{CO2} dominated). As the exoplanets simulated here have limited areas where surface temperatures are above 273 K (and some have no surface temperatures above 273 K), subsequent research could investigate the parameter space with warmer exoplanets to determine how surface \ce{O3} concentrations are affected. Examples of other 3D models that could investigate this chemical phenomenon are LMD-G \citep{2022CliPa..18.2421Y}, ROCKE-3D \citep{2017ApJS..231...12W}, and the Unified Model \citep{2017A&A...601A.120B}. Once a chemical scheme is implemented, the LFRic-Atmosphere model could also be used \citep{2023GMD....16.5601S}. 

The classical HZ depends on whether a planet can sustain surface liquid water, but the notion of a Habitable Zone for Complex Life \citep{2019ApJ...878...19S} is influenced by the presence of toxic gases. So far, the toxic gases that have been suggested are carbon monoxide (CO) and relatively high concentrations of \ce{CO2} \citep{2019ApJ...878...19S}, as well as \ce{N2} at high pressure \citep[e.g., $ > 2$ bar;][]{2020NatSR..10.7432R}. These molecules impede biochemical processes that may be unique to organisms on Earth. Resulting from its reactivity as a strong oxidant, \ce{O3} could pose a more significant threat to extraterrestrial life. Therefore, based on our modelling work and the properties of \ce{O3}, we recommend that \ce{O3} should now be added to the list of molecules that can influence the Habitable Zone for Complex Life.
 
\section{Conclusions}
\label{Conclusions section}

This work used WACCM6 to simulate the climate of two exoplanets: TRAPPIST-1 e and Proxima Centauri b. For each exoplanet, we considered \ce{O2} mixing ratios between 0.1\% PAL and 100\% PAL. Additionally, two different stellar spectra were used for the TRAPPIST-1 e cases to investigate the effect on surface \ce{O3} due to their large differences in the strength of incoming UV radiation. In multiple simulations, surface concentrations of \ce{O3} exceed 40 ppbv, with maximum time-averaged concentrations reaching up to 2200 ppbv in the TP-1 e P19 PI case. Such concentrations are harmful to life on Earth and may be potentially fatal through oxidative stress. In these simulated atmospheres, \ce{O3} exists not as a pollutant, but as a consequence of the planetary atmospheric conditions, such as the 1,000 hPa surface pressure, the incoming UV strength and shape, and the \ce{O2} number density vertical profile. Our work suggests the potential presence of toxic \ce{O3} concentrations should be included when evaluating the habitability of an exoplanet.

The simulations examined in this exploratory work represent a small proportion of the parameter space in which atmospheres may form relatively high \ce{O3} concentrations at the surface. Different planetary rotation rates, topography, atmospheric pressures, total irradiation and UV irradiation environments, as well as various chemical fluxes from the surface to the atmosphere, should all be explored. Upcoming work should consider the potential presence of high surface concentrations of \ce{O3} when simulating oxygenated atmospheres. If \ce{O3} is detected in any future observations of terrestrial exoplanet atmospheres, ascertaining the \ce{O3} surface concentration should be incorporated into frameworks that aim to determine planetary habitability and decide on the most promising targets for follow-up observations \citep[see e.g., ][]{2020AJ....159...55T, 2021AsBio..21.1017M, 2021EPJST.230.2207S}. In practise, this will require a combination of planetary modelling, transmission and direct imaging spectra, as well as precise knowledge of the UV irradiation environment of the atmosphere. 3D chemistry-climate models are essential for understanding how transport can create areas with comparatively lower and thus safer \ce{O3} concentrations. Just as on Earth, the entire surface does not need to be hospitable for life to flourish.

\section*{Acknowledgments}

We thank the two reviewers for their comments and helpful feedback which helped to improve the manuscript.

G.J.C. acknowledges the studentship funded by the Science and Technology Facilities Council of the United Kingdom (STFC; grant number ST/T506230/1). C.W. acknowledges financial support from the University of Leeds and from the Science and Technology Facilities Council (grant numbers ST/X001016/1 and MR/T040726/1). F.S.-M. acknowledges financial support from the University of Leeds and from the Science and Technology Facilities Council (grant number MR/T040726/1). This work was undertaken on ARC4, part of the High Performance Computing facilities at the University of Leeds, UK.

\bibliography{references}{}

\begin{thebibliography}{}
\expandafter\ifx\csname natexlab\endcsname\relax\def\natexlab#1{#1}\fi
\providecommand{\url}[1]{\href{#1}{#1}}
\providecommand{\dodoi}[1]{doi:~\href{http://doi.org/#1}{\nolinkurl{#1}}}
\providecommand{\doeprint}[1]{\href{http://ascl.net/#1}{\nolinkurl{http://ascl.net/#1}}}
\providecommand{\doarXiv}[1]{\href{https://arxiv.org/abs/#1}{\nolinkurl{https://arxiv.org/abs/#1}}}

\bibitem[{Adams {et~al.}(1974)Adams, Titus, \& McCracken}]{adams1974depth}
Adams, M.~S., Titus, J., \& McCracken, M. 1974, Limnology and oceanography, 19,
  377

\bibitem[{{Agol} {et~al.}(2021){Agol}, {Dorn}, {Grimm}, {Turbet}, {Ducrot},
  {Delrez}, {Gillon}, {Demory}, {Burdanov}, {Barkaoui}, {Benkhaldoun},
  {Bolmont}, {Burgasser}, {Carey}, {de Wit}, {Fabrycky}, {Foreman-Mackey},
  {Haldemann}, {Hernandez}, {Ingalls}, {Jehin}, {Langford}, {Leconte},
  {Lederer}, {Luger}, {Malhotra}, {Meadows}, {Morris}, {Pozuelos}, {Queloz},
  {Raymond}, {Selsis}, {Sestovic}, {Triaud}, \& {Van
  Grootel}}]{2021PSJ.....2....1A}
{Agol}, E., {Dorn}, C., {Grimm}, S.~L., {et~al.} 2021, \psj, 2, 1,
  \dodoi{10.3847/PSJ/abd022}

\bibitem[{Ainsworth {et~al.}(2012)Ainsworth, Yendrek, Sitch, Collins, \&
  Emberson}]{ainsworth2012effects}
Ainsworth, E.~A., Yendrek, C.~R., Sitch, S., Collins, W.~J., \& Emberson, L.~D.
  2012, Annual review of plant biology, 63, 637

\bibitem[{{Alados} {et~al.}(1996){Alados}, {Foyo-Moreno}, \&
  {Alados-Arboledas}}]{1996AgFM...78..121A}
{Alados}, I., {Foyo-Moreno}, I., \& {Alados-Arboledas}, L. 1996, Agricultural
  and Forest Meteorology, 78, 121, \dodoi{10.1016/0168-1923(95)02245-7}

\bibitem[{{Arney} {et~al.}(2016){Arney}, {Domagal-Goldman}, {Meadows}, {Wolf},
  {Schwieterman}, {Charnay}, {Claire}, {H{\'e}brard}, \&
  {Trainer}}]{2016AsBio..16..873A}
{Arney}, G., {Domagal-Goldman}, S.~D., {Meadows}, V.~S., {et~al.} 2016,
  Astrobiology, 16, 873, \dodoi{10.1089/ast.2015.1422}

\bibitem[{{Arney}(2019)}]{2019ApJ...873L...7A}
{Arney}, G.~N. 2019, \apjl, 873, L7, \dodoi{10.3847/2041-8213/ab0651}

\bibitem[{Arrigo {et~al.}(2012)Arrigo, Perovich, Pickart, Brown, Van~Dijken,
  Lowry, Mills, Palmer, Balch, Bahr, {et~al.}}]{arrigo2012massive}
Arrigo, K.~R., Perovich, D.~K., Pickart, R.~S., {et~al.} 2012, Science, 336,
  1408

\bibitem[{{Atkinson}(2000)}]{2000AtmEn..34.2063A}
{Atkinson}, R. 2000, Atmospheric Environment, 34, 2063,
  \dodoi{10.1016/S1352-2310(99)00460-4}

\bibitem[{{Avnery} {et~al.}(2011){Avnery}, {Mauzerall}, {Liu}, \&
  {Horowitz}}]{2011AtmEn..45.2284A}
{Avnery}, S., {Mauzerall}, D.~L., {Liu}, J., \& {Horowitz}, L.~W. 2011,
  Atmospheric Environment, 45, 2284, \dodoi{10.1016/j.atmosenv.2010.11.045}

\bibitem[{Barten {et~al.}(2023)Barten, Ganzeveld, Steeneveld, Blomquist, Angot,
  Archer, Bariteau, Beck, Boyer, Von~der Gathen, {et~al.}}]{barten2023low}
Barten, J.~G., Ganzeveld, L.~N., Steeneveld, G.-J., {et~al.} 2023, Elem Sci
  Anth, 11, 00086

\bibitem[{{Barth} {et~al.}(2024){Barth}, {St{\"u}eken}, {Helling},
  {Schwieterman}, \& {Telling}}]{2024arXiv240213682B}
{Barth}, P., {St{\"u}eken}, E.~E., {Helling}, C., {Schwieterman}, E.~W., \&
  {Telling}, J. 2024, arXiv e-prints, arXiv:2402.13682,
  \dodoi{10.48550/arXiv.2402.13682}

\bibitem[{Behrendt {et~al.}(2020)Behrendt, Trampe, Nord, Nguyen, K{\"u}hl,
  Lonco, Nyarko, Dhinojwala, Hershey, \& Barton}]{behrendt2020life}
Behrendt, L., Trampe, E.~L., Nord, N.~B., {et~al.} 2020, Environmental
  microbiology, 22, 952

\bibitem[{Bell {et~al.}(2006)Bell, Peng, \& Dominici}]{bell2006exposure}
Bell, M.~L., Peng, R.~D., \& Dominici, F. 2006, Environmental health
  perspectives, 114, 532

\bibitem[{Borlongan {et~al.}(2017)Borlongan, Nishihara, Shimada, \&
  Terada}]{borlongan2017photosynthetic}
Borlongan, I.~A., Nishihara, G.~N., Shimada, S., \& Terada, R. 2017, Journal of
  applied phycology, 29, 3077

\bibitem[{{Bortkovskii}(2012)}]{2012IzAOP..48..538B}
{Bortkovskii}, R.~S. 2012, Izvestiya Atmospheric and Oceanic Physics, 48, 538,
  \dodoi{10.1134/S0001433812040044}

\bibitem[{{Boutle} {et~al.}(2017){Boutle}, {Mayne}, {Drummond}, {Manners},
  {Goyal}, {Hugo Lambert}, {Acreman}, \& {Earnshaw}}]{2017A&A...601A.120B}
{Boutle}, I.~A., {Mayne}, N.~J., {Drummond}, B., {et~al.} 2017, \aap, 601,
  A120, \dodoi{10.1051/0004-6361/201630020}

\bibitem[{{Braam} {et~al.}(2023){Braam}, {Palmer}, {Decin}, {Cohen}, \&
  {Mayne}}]{2023MNRAS.526..263B}
{Braam}, M., {Palmer}, P.~I., {Decin}, L., {Cohen}, M., \& {Mayne}, N.~J. 2023,
  \mnras, 526, 263, \dodoi{10.1093/mnras/stad2704}

\bibitem[{{Braam} {et~al.}(2022){Braam}, {Palmer}, {Decin}, {Ridgway},
  {Zamyatina}, {Mayne}, {Sergeev}, \& {Abraham}}]{2022MNRAS.517.2383B}
{Braam}, M., {Palmer}, P.~I., {Decin}, L., {et~al.} 2022, \mnras, 517, 2383,
  \dodoi{10.1093/mnras/stac2722}

\bibitem[{{Brasseur} \& {Solomon}(2005)}]{2005ama..book.....B}
{Brasseur}, G.~P., \& {Solomon}, S. 2005, {Aeronomy of the Middle Atmosphere:
  Chemistry and Physics of the Stratosphere and Mesosphere} (Springer)

\bibitem[{{Brugger} {et~al.}(2016){Brugger}, {Mousis}, {Deleuil}, \&
  {Lunine}}]{2016ApJ...831L..16B}
{Brugger}, B., {Mousis}, O., {Deleuil}, M., \& {Lunine}, J.~I. 2016, \apjl,
  831, L16, \dodoi{10.3847/2041-8205/831/2/L16}

\bibitem[{Butchart(2014)}]{butchart2014brewer}
Butchart, N. 2014, Reviews of geophysics, 52, 157

\bibitem[{Bytnerowicz {et~al.}(1993)Bytnerowicz, Manning, Grosjean,
  Chmielewski, Dmuchowski, Grodzinska, \& Godzik}]{bytnerowicz1993detecting}
Bytnerowicz, A., Manning, W.~J., Grosjean, D., {et~al.} 1993, Environmental
  Pollution, 80, 301

\bibitem[{{Campbell} {et~al.}(2007){Campbell}, {McKenzie}, {Kerville}, \&
  {Bit{\'e}}}]{2007ECSS...73..551C}
{Campbell}, S.~J., {McKenzie}, L.~J., {Kerville}, S.~P., \& {Bit{\'e}}, J.~S.
  2007, Estuarine Coastal and Shelf Science, 73, 551,
  \dodoi{10.1016/j.ecss.2007.02.014}

\bibitem[{{Carone} {et~al.}(2018){Carone}, {Keppens}, {Decin}, \&
  {Henning}}]{2018MNRAS.473.4672C}
{Carone}, L., {Keppens}, R., {Decin}, L., \& {Henning}, T. 2018, \mnras, 473,
  4672, \dodoi{10.1093/mnras/stx2732}

\bibitem[{{Catling} \& {Zahnle}(2020)}]{2020SciA....6.1420C}
{Catling}, D.~C., \& {Zahnle}, K.~J. 2020, Science Advances, 6, eaax1420,
  \dodoi{10.1126/sciadv.aax1420}

\bibitem[{Catling {et~al.}(2001)Catling, Zahnle, \&
  McKay}]{catling2001biogenic}
Catling, D.~C., Zahnle, K.~J., \& McKay, C.~P. 2001, Science, 293, 839

\bibitem[{{Chameides} {et~al.}(1988){Chameides}, {Lindsay}, {Richardson}, \&
  {Kiang}}]{1988Sci...241.1473C}
{Chameides}, W.~L., {Lindsay}, R.~W., {Richardson}, J., \& {Kiang}, C.~S. 1988,
  Science, 241, 1473, \dodoi{10.1126/science.3420404}

\bibitem[{{Chen} {et~al.}(2019){Chen}, {Wolf}, {Zhan}, \&
  {Horton}}]{2019ApJ...886...16C}
{Chen}, H., {Wolf}, E.~T., {Zhan}, Z., \& {Horton}, D.~E. 2019, \apj, 886, 16,
  \dodoi{10.3847/1538-4357/ab4f7e}

\bibitem[{{Chen} {et~al.}(2021){Chen}, {Zhan}, {Youngblood}, {Wolf},
  {Feinstein}, \& {Horton}}]{2021NatAs...5..298C}
{Chen}, H., {Zhan}, Z., {Youngblood}, A., {et~al.} 2021, Nature Astronomy, 5,
  298, \dodoi{10.1038/s41550-020-01264-1}

\bibitem[{{Claudi} {et~al.}(2020){Claudi}, {Alei}, {Battistuzzi}, {Cocola},
  {Erculiani}, {Pozzer}, {Salasnich}, {Simionato}, {Squicciarini}, {Poletto},
  \& {La Rocca}}]{2020Life...11...10C}
{Claudi}, R., {Alei}, E., {Battistuzzi}, M., {et~al.} 2020, Life, 11, 10,
  \dodoi{10.3390/life11010010}

\bibitem[{{Clement Kinney} {et~al.}(2020){Clement Kinney}, {Maslowski},
  {Osinski}, {Jin}, {Frants}, {Jeffery}, \& {Lee}}]{2020JGRC..12516211C}
{Clement Kinney}, J., {Maslowski}, W., {Osinski}, R., {et~al.} 2020, Journal of
  Geophysical Research (Oceans), 125, e2020JC016211,
  \dodoi{10.1029/2020JC016211}

\bibitem[{{Clifton} {et~al.}(2020){Clifton}, {Fiore}, {Massman}, {Baublitz},
  {Coyle}, {Emberson}, {Fares}, {Farmer}, {Gentine}, {Gerosa}, {Guenther},
  {Helmig}, {Lombardozzi}, {Munger}, {Patton}, {Pusede}, {Schwede}, {Silva},
  {S{\"o}rgel}, {Steiner}, \& {Tai}}]{2020RvGeo..5800670C}
{Clifton}, O.~E., {Fiore}, A.~M., {Massman}, W.~J., {et~al.} 2020, Reviews of
  Geophysics, 58, e2019RG000670, \dodoi{10.1029/2019RG000670}

\bibitem[{{Colose} {et~al.}(2019){Colose}, {Del Genio}, \&
  {Way}}]{2019ApJ...884..138C}
{Colose}, C.~M., {Del Genio}, A.~D., \& {Way}, M.~J. 2019, \apj, 884, 138,
  \dodoi{10.3847/1538-4357/ab4131}

\bibitem[{Cooke {et~al.}(2023)Cooke, Marsh, Walsh, \&
  Youngblood}]{cooke2023degenerate}
Cooke, G., Marsh, D., Walsh, C., \& Youngblood, A. 2023, arXiv preprint
  arXiv:2309.15239

\bibitem[{{Cooke} {et~al.}(2022){Cooke}, {Marsh}, {Walsh}, {Black}, \&
  {Lamarque}}]{2022RSOS....911165C}
{Cooke}, G.~J., {Marsh}, D.~R., {Walsh}, C., {Black}, B., \& {Lamarque}, J.~F.
  2022, Royal Society Open Science, 9, 211165, \dodoi{10.1098/rsos.211165}

\bibitem[{{Cooke} {et~al.}(2023){Cooke}, {Marsh}, {Walsh}, {Rugheimer}, \&
  {Villanueva}}]{2023MNRAS.518..206C}
{Cooke}, G.~J., {Marsh}, D.~R., {Walsh}, C., {Rugheimer}, S., \& {Villanueva},
  G.~L. 2023, \mnras, 518, 206, \dodoi{10.1093/mnras/stac2604}

\bibitem[{{Covone} {et~al.}(2021){Covone}, {Ienco}, {Cacciapuoti}, \&
  {Inno}}]{2021MNRAS.505.3329C}
{Covone}, G., {Ienco}, R.~M., {Cacciapuoti}, L., \& {Inno}, L. 2021, \mnras,
  505, 3329, \dodoi{10.1093/mnras/stab1357}

\bibitem[{{Del Genio} {et~al.}(2019){Del Genio}, {Way}, {Amundsen}, {Aleinov},
  {Kelley}, {Kiang}, \& {Clune}}]{2019AsBio..19...99D}
{Del Genio}, A.~D., {Way}, M.~J., {Amundsen}, D.~S., {et~al.} 2019,
  Astrobiology, 19, 99, \dodoi{10.1089/ast.2017.1760}

\bibitem[{{Delille} {et~al.}(2007){Delille}, {Jourdain}, {Borges}, {Tison}, \&
  {Delille}}]{2007LimOc..52.1367D}
{Delille}, B., {Jourdain}, B., {Borges}, A.~V., {Tison}, J.-L., \& {Delille},
  D. 2007, Limnology and Oceanography, 52, 1367,
  \dodoi{10.4319/lo.2007.52.4.1367}

\bibitem[{{D{\'e}mares} {et~al.}(2022){D{\'e}mares}, {Gibert}, {Creusot},
  {Lapeyre}, \& {Proffit}}]{2022ScTEn.827o4342D}
{D{\'e}mares}, F., {Gibert}, L., {Creusot}, P., {Lapeyre}, B., \& {Proffit}, M.
  2022, Science of the Total Environment, 827, 154342,
  \dodoi{10.1016/j.scitotenv.2022.154342}

\bibitem[{{Des Marais} {et~al.}(2002){Des Marais}, {Harwit}, {Jucks},
  {Kasting}, {Lin}, {Lunine}, {Schneider}, {Seager}, {Traub}, \&
  {Woolf}}]{2002AsBio...2..153D}
{Des Marais}, D.~J., {Harwit}, M.~O., {Jucks}, K.~W., {et~al.} 2002,
  Astrobiology, 2, 153, \dodoi{10.1089/15311070260192246}

\bibitem[{{Domagal-Goldman} {et~al.}(2014){Domagal-Goldman}, {Segura},
  {Claire}, {Robinson}, \& {Meadows}}]{2014ApJ...792...90D}
{Domagal-Goldman}, S.~D., {Segura}, A., {Claire}, M.~W., {Robinson}, T.~D., \&
  {Meadows}, V.~S. 2014, \apj, 792, 90, \dodoi{10.1088/0004-637X/792/2/90}

\bibitem[{{Duffy} {et~al.}(2023){Duffy}, {Canchon}, {Haworth}, {Gillen},
  {Chitnavis}, \& {Mullineaux}}]{2023MNRAS.526.2265D}
{Duffy}, C. D.~P., {Canchon}, G., {Haworth}, T.~J., {et~al.} 2023, \mnras, 526,
  2265, \dodoi{10.1093/mnras/stad2823}

\bibitem[{Egorova {et~al.}(2015)Egorova, Voblikova, Sabitova, Tkachenko,
  Tkachenko, \& Lunin}]{egorova2015ozone}
Egorova, G., Voblikova, V., Sabitova, L., {et~al.} 2015, Moscow University
  Chemistry Bulletin, 70, 207

\bibitem[{{El-Madany} {et~al.}(2017){El-Madany}, {Niklasch}, \&
  {Klemm}}]{2017Atmos...8..175E}
{El-Madany}, T., {Niklasch}, K., \& {Klemm}, O. 2017, Atmosphere, 8, 175,
  \dodoi{10.3390/atmos8090175}

\bibitem[{{Emmons} {et~al.}(2020){Emmons}, {Schwantes}, {Orlando}, {Tyndall},
  {Kinnison}, {Lamarque}, {Marsh}, {Mills}, {Tilmes}, {Bardeen}, {Buchholz},
  {Conley}, {Gettelman}, {Garcia}, {Simpson}, {Blake}, {Meinardi}, \&
  {P{\'e}tron}}]{2020JAMES..1201882E}
{Emmons}, L.~K., {Schwantes}, R.~H., {Orlando}, J.~J., {et~al.} 2020, Journal
  of Advances in Modeling Earth Systems, 12, e2019MS001882,
  \dodoi{10.1029/2019MS001882}

\bibitem[{Epelle {et~al.}(2023)Epelle, Macfarlane, Cusack, Burns, Okolie,
  Mackay, Rateb, \& Yaseen}]{epelle2023ozone}
Epelle, E.~I., Macfarlane, A., Cusack, M., {et~al.} 2023, Chemical Engineering
  Journal, 454, 140188

\bibitem[{Epelle {et~al.}(2022)Epelle, Macfarlane, Cusack, Burns, Thissera,
  Mackay, Rateb, \& Yaseen}]{epelle2022bacterial}
---. 2022, Journal of Microbiological Methods, 194, 106431

\bibitem[{Erickson \& Ortega(2006)}]{erickson2006inactivation}
Erickson, M.~C., \& Ortega, Y.~R. 2006, Journal of food protection, 69, 2786

\bibitem[{{Faria} {et~al.}(2022){Faria}, {Su{\'a}rez Mascare{\~n}o},
  {Figueira}, {Silva}, {Damasso}, {Demangeon}, {Pepe}, {Santos}, {Rebolo},
  {Cristiani}, {Adibekyan}, {Alibert}, {Allart}, {Barros}, {Cabral},
  {D'Odorico}, {Di Marcantonio}, {Dumusque}, {Ehrenreich}, {Gonz{\'a}lez
  Hern{\'a}ndez}, {Hara}, {Lillo-Box}, {Lo Curto}, {Lovis}, {Martins},
  {M{\'e}gevand}, {Mehner}, {Micela}, {Molaro}, {Nunes}, {Pall{\'e}},
  {Poretti}, {Sousa}, {Sozzetti}, {Tabernero}, {Udry}, \& {Zapatero
  Osorio}}]{2022A&A...658A.115F}
{Faria}, J.~P., {Su{\'a}rez Mascare{\~n}o}, A., {Figueira}, P., {et~al.} 2022,
  \aap, 658, A115, \dodoi{10.1051/0004-6361/202142337}

\bibitem[{{Fauchez} {et~al.}(2019){Fauchez}, {Turbet}, {Villanueva}, {Wolf},
  {Arney}, {Kopparapu}, {Lincowski}, {Mandell}, {de Wit}, {Pidhorodetska},
  {Domagal-Goldman}, \& {Stevenson}}]{2019ApJ...887..194F}
{Fauchez}, T.~J., {Turbet}, M., {Villanueva}, G.~L., {et~al.} 2019, \apj, 887,
  194, \dodoi{10.3847/1538-4357/ab5862}

\bibitem[{{Fauchez} {et~al.}(2020){Fauchez}, {Turbet}, {Wolf}, {Boutle}, {Way},
  {Del Genio}, {Mayne}, {Tsigaridis}, {Kopparapu}, {Yang}, {Forget}, {Mandell},
  \& {Domagal Goldman}}]{2020GMD....13..707F}
{Fauchez}, T.~J., {Turbet}, M., {Wolf}, E.~T., {et~al.} 2020, Geoscientific
  Model Development, 13, 707, \dodoi{10.5194/gmd-13-707-2020}

\bibitem[{Feng {et~al.}(2022)Feng, Xu, Kobayashi, Dai, Zhang, Agathokleous,
  Calatayud, Paoletti, Mukherjee, Agrawal, {et~al.}}]{feng2022ozone}
Feng, Z., Xu, Y., Kobayashi, K., {et~al.} 2022, Nature Food, 3, 47

\bibitem[{{Field} {et~al.}(1998){Field}, {Behrenfeld}, {Randerson}, \&
  {Falkowski}}]{1998Sci...281..237F}
{Field}, C.~B., {Behrenfeld}, M.~J., {Randerson}, J.~T., \& {Falkowski}, P.
  1998, Science, 281, 237, \dodoi{10.1126/science.281.5374.237}

\bibitem[{Fontes {et~al.}(2012)Fontes, Cattani~Heimbecker, de~Souza~Brito,
  Costa, van~der Heijden, Levin, \& Rasslan}]{fontes2012effect}
Fontes, B., Cattani~Heimbecker, A.~M., de~Souza~Brito, G., {et~al.} 2012, BMC
  infectious diseases, 12, 1

\bibitem[{{Fowler} {et~al.}(2009){Fowler}, {Pilegaard}, {Sutton}, {Ambus},
  {Raivonen}, {Duyzer}, {Simpson}, {Fagerli}, {Fuzzi}, {Schjoerring},
  {Granier}, {Neftel}, {Isaksen}, {Laj}, {Maione}, {Monks}, {Burkhardt},
  {Daemmgen}, {Neirynck}, {Personne}, {Wichink-Kruit}, {Butterbach-Bahl},
  {Flechard}, {Tuovinen}, {Coyle}, {Gerosa}, {Loubet}, {Altimir}, {Gruenhage},
  {Ammann}, {Cieslik}, {Paoletti}, {Mikkelsen}, {Ro-Poulsen}, {Cellier},
  {Cape}, {Horv{\'a}th}, {Loreto}, {Niinemets}, {Palmer}, {Rinne}, {Misztal},
  {Nemitz}, {Nilsson}, {Pryor}, {Gallagher}, {Vesala}, {Skiba},
  {Br{\"u}ggemann}, {Zechmeister-Boltenstern}, {Williams}, {O'Dowd},
  {Facchini}, {de Leeuw}, {Flossman}, {Chaumerliac}, \&
  {Erisman}}]{2009AtmEn..43.5193F}
{Fowler}, D., {Pilegaard}, K., {Sutton}, M.~A., {et~al.} 2009, Atmospheric
  Environment, 43, 5193, \dodoi{10.1016/j.atmosenv.2009.07.068}

\bibitem[{{Fowler} {et~al.}(2023){Fowler}, {Haffert}, {van Kooten}, {Landman},
  {Bidot}, {Hours}, {N'Diaye}, {Absil}, {Altinier}, {Baudoz}, {Belikov},
  {Bonse}, {Bott}, {Brandl}, {Carlotti}, {Casewell}, {Choquet}, {Cowan},
  {Desai}, {Doelman}, {Fogarty}, {Gebhard}, {Gutierrez}, {Guyon},
  {Herscovici-Schiller}, {Juanola-Parramon}, {Kenworthy}, {Kleisioti}, {Konig},
  {Krasteva}, {Laginja}, {Leboulleux}, {Mazoyer}, {Millar-Blanchaer},
  {Mouillet}, {Por}, {Pueyo}, {Snik}, {van Dam}, {van Gorkom}, \&
  {Vaughan}}]{2023arXiv230900725F}
{Fowler}, J., {Haffert}, S.~Y., {van Kooten}, M. A.~M., {et~al.} 2023, arXiv
  e-prints, arXiv:2309.00725, \dodoi{10.48550/arXiv.2309.00725}

\bibitem[{{France} {et~al.}(2016){France}, {Loyd}, {Youngblood}, {Brown},
  {Schneider}, {Hawley}, {Froning}, {Linsky}, {Roberge}, {Buccino},
  {Davenport}, {Fontenla}, {Kaltenegger}, {Kowalski}, {Mauas}, {Miguel},
  {Redfield}, {Rugheimer}, {Tian}, {Vieytes}, {Walkowicz}, \&
  {Weisenburger}}]{2016ApJ...820...89F}
{France}, K., {Loyd}, R.~O.~P., {Youngblood}, A., {et~al.} 2016, \apj, 820, 89,
  \dodoi{10.3847/0004-637X/820/2/89}

\bibitem[{{Froning} {et~al.}(2019){Froning}, {Kowalski}, {France}, {Loyd},
  {Schneider}, {Youngblood}, {Wilson}, {Brown}, {Berta-Thompson}, {Pineda},
  {Linsky}, {Rugheimer}, \& {Miguel}}]{2019ApJ...871L..26F}
{Froning}, C.~S., {Kowalski}, A., {France}, K., {et~al.} 2019, \apjl, 871, L26,
  \dodoi{10.3847/2041-8213/aaffcd}

\bibitem[{{Fu} {et~al.}(2019){Fu}, {Solomon}, {Pahlavan}, \&
  {Lin}}]{2019ERL....14k4026F}
{Fu}, Q., {Solomon}, S., {Pahlavan}, H.~A., \& {Lin}, P. 2019, Environmental
  Research Letters, 14, 114026, \dodoi{10.1088/1748-9326/ab4de7}

\bibitem[{{Gaia Collaboration} {et~al.}(2016){Gaia Collaboration}, {Brown},
  {Vallenari}, {Prusti}, {de Bruijne}, {Mignard}, {Drimmel}, {Babusiaux},
  {Bailer-Jones}, {Bastian}, {Biermann}, {Evans}, {Eyer}, {Jansen}, {Jordi},
  {Katz}, {Klioner}, {Lammers}, {Lindegren}, {Luri}, {O'Mullane}, {Panem},
  {Pourbaix}, {Randich}, {Sartoretti}, {Siddiqui}, {Soubiran}, {Valette}, {van
  Leeuwen}, {Walton}, {Aerts}, {Arenou}, {Cropper}, {H{\o}g}, {Lattanzi},
  {Grebel}, {Holland}, {Huc}, {Passot}, {Perryman}, {Bramante}, {Cacciari},
  {Casta{\~n}eda}, {Chaoul}, {Cheek}, {De Angeli}, {Fabricius}, {Guerra},
  {Hern{\'a}ndez}, {Jean-Antoine-Piccolo}, {Masana}, {Messineo}, {Mowlavi},
  {Nienartowicz}, {Ord{\'o}{\~n}ez-Blanco}, {Panuzzo}, {Portell}, {Richards},
  {Riello}, {Seabroke}, {Tanga}, {Th{\'e}venin}, {Torra}, {Els},
  {Gracia-Abril}, {Comoretto}, {Garcia-Reinaldos}, {Lock}, {Mercier},
  {Altmann}, {Andrae}, {Astraatmadja}, {Bellas-Velidis}, {Benson}, {Berthier},
  {Blomme}, {Busso}, {Carry}, {Cellino}, {Clementini}, {Cowell}, {Creevey},
  {Cuypers}, {Davidson}, {De Ridder}, {de Torres}, {Delchambre}, {Dell'Oro},
  {Ducourant}, {Fr{\'e}mat}, {Garc{\'\i}a-Torres}, {Gosset}, {Halbwachs},
  {Hambly}, {Harrison}, {Hauser}, {Hestroffer}, {Hodgkin}, {Huckle}, {Hutton},
  {Jasniewicz}, {Jordan}, {Kontizas}, {Korn}, {Lanzafame}, {Manteiga},
  {Moitinho}, {Muinonen}, {Osinde}, {Pancino}, {Pauwels}, {Petit},
  {Recio-Blanco}, {Robin}, {Sarro}, {Siopis}, {Smith}, {Smith}, {Sozzetti},
  {Thuillot}, {van Reeven}, {Viala}, {Abbas}, {Abreu Aramburu}, {Accart},
  {Aguado}, {Allan}, {Allasia}, {Altavilla}, {{\'A}lvarez}, {Alves},
  {Anderson}, {Andrei}, {Anglada Varela}, {Antiche}, {Antoja}, {Ant{\'o}n},
  {Arcay}, {Bach}, {Baker}, {Balaguer-N{\'u}{\~n}ez}, {Barache}, {Barata},
  {Barbier}, {Barblan}, {Barrado y Navascu{\'e}s}, {Barros}, {Barstow},
  {Becciani}, {Bellazzini}, {Bello Garc{\'\i}a}, {Belokurov}, {Bendjoya},
  {Berihuete}, {Bianchi}, {Bienaym{\'e}}, {Billebaud}, {Blagorodnova},
  {Blanco-Cuaresma}, {Boch}, {Bombrun}, {Borrachero}, {Bouquillon}, {Bourda},
  {Bouy}, {Bragaglia}, {Breddels}, {Brouillet}, {Br{\"u}semeister},
  {Bucciarelli}, {Burgess}, {Burgon}, {Burlacu}, {Busonero}, {Buzzi}, {Caffau},
  {Cambras}, {Campbell}, {Cancelliere}, {Cantat-Gaudin}, {Carlucci},
  {Carrasco}, {Castellani}, {Charlot}, {Charnas}, {Chiavassa}, {Clotet},
  {Cocozza}, {Collins}, {Costigan}, {Crifo}, {Cross}, {Crosta}, {Crowley},
  {Dafonte}, {Damerdji}, {Dapergolas}, {David}, {David}, {De Cat}, {de Felice},
  {de Laverny}, {De Luise}, {De March}, {de Martino}, {de Souza}, {Debosscher},
  {del Pozo}, {Delbo}, {Delgado}, {Delgado}, {Di Matteo}, {Diakite},
  {Distefano}, {Dolding}, {Dos Anjos}, {Drazinos}, {Duran}, {Dzigan},
  {Edvardsson}, {Enke}, {Evans}, {Eynard Bontemps}, {Fabre}, {Fabrizio},
  {Faigler}, {Falc{\~a}o}, {Farr{\`a}s Casas}, {Federici}, {Fedorets},
  {Fern{\'a}ndez-Hern{\'a}ndez}, {Fernique}, {Fienga}, {Figueras}, {Filippi},
  {Findeisen}, {Fonti}, {Fouesneau}, {Fraile}, {Fraser}, {Fuchs}, {Gai},
  {Galleti}, {Galluccio}, {Garabato}, {Garc{\'\i}a-Sedano}, {Garofalo},
  {Garralda}, {Gavras}, {Gerssen}, {Geyer}, {Gilmore}, {Girona}, {Giuffrida},
  {Gomes}, {Gonz{\'a}lez-Marcos}, {Gonz{\'a}lez-N{\'u}{\~n}ez},
  {Gonz{\'a}lez-Vidal}, {Granvik}, {Guerrier}, {Guillout}, {Guiraud},
  {G{\'u}rpide}, {Guti{\'e}rrez-S{\'a}nchez}, {Guy}, {Haigron},
  {Hatzidimitriou}, {Haywood}, {Heiter}, {Helmi}, {Hobbs}, {Hofmann}, {Holl},
  {Holland}, {Hunt}, {Hypki}, {Icardi}, {Irwin}, {Jevardat de Fombelle},
  {Jofr{\'e}}, {Jonker}, {Jorissen}, {Julbe}, {Karampelas}, {Kochoska},
  {Kohley}, {Kolenberg}, {Kontizas}, {Koposov}, {Kordopatis}, {Koubsky},
  {Krone-Martins}, {Kudryashova}, {Kull}, {Bachchan}, {Lacoste-Seris}, {Lanza},
  {Lavigne}, {Le Poncin-Lafitte}, {Lebreton}, {Lebzelter}, {Leccia}, {Leclerc},
  {Lecoeur-Taibi}, {Lemaitre}, {Lenhardt}, {Leroux}, {Liao}, {Licata},
  {Lindstr{\o}m}, {Lister}, {Livanou}, {Lobel}, {L{\"o}ffler}, {L{\'o}pez},
  {Lorenz}, {MacDonald}, {Magalh{\~a}es Fernandes}, {Managau}, {Mann},
  {Mantelet}, {Marchal}, {Marchant}, {Marconi}, {Marinoni}, {Marrese},
  {Marschalk{\'o}}, {Marshall}, {Mart{\'\i}n-Fleitas}, {Martino}, {Mary},
  {Matijevi{\v{c}}}, {Mazeh}, {McMillan}, {Messina}, {Michalik}, {Millar},
  {Miranda}, {Molina}, {Molinaro}, {Molinaro}, {Moln{\'a}r}, {Moniez},
  {Montegriffo}, {Mor}, {Mora}, {Morbidelli}, {Morel}, {Morgenthaler},
  {Morris}, {Mulone}, {Muraveva}, {Musella}, {Narbonne}, {Nelemans},
  {Nicastro}, {Noval}, {Ord{\'e}novic}, {Ordieres-Mer{\'e}}, {Osborne},
  {Pagani}, {Pagano}, {Pailler}, {Palacin}, {Palaversa}, {Parsons}, {Pecoraro},
  {Pedrosa}, {Pentik{\"a}inen}, {Pichon}, {Piersimoni}, {Pineau}, {Plachy},
  {Plum}, {Poujoulet}, {Pr{\v{s}}a}, {Pulone}, {Ragaini}, {Rago}, {Rambaux},
  {Ramos-Lerate}, {Ranalli}, {Rauw}, {Read}, {Regibo}, {Reyl{\'e}}, {Ribeiro},
  {Rimoldini}, {Ripepi}, {Riva}, {Rixon}, {Roelens}, {Romero-G{\'o}mez},
  {Rowell}, {Royer}, {Ruiz-Dern}, {Sadowski}, {Sagrist{\`a} Sell{\'e}s},
  {Sahlmann}, {Salgado}, {Salguero}, {Sarasso}, {Savietto}, {Schultheis},
  {Sciacca}, {Segol}, {Segovia}, {Segransan}, {Shih}, {Smareglia}, {Smart},
  {Solano}, {Solitro}, {Sordo}, {Soria Nieto}, {Souchay}, {Spagna}, {Spoto},
  {Stampa}, {Steele}, {Steidelm{\"u}ller}, {Stephenson}, {Stoev}, {Suess},
  {S{\"u}veges}, {Surdej}, {Szabados}, {Szegedi-Elek}, {Tapiador}, {Taris},
  {Tauran}, {Taylor}, {Teixeira}, {Terrett}, {Tingley}, {Trager}, {Turon},
  {Ulla}, {Utrilla}, {Valentini}, {van Elteren}, {Van Hemelryck}, {van
  Leeuwen}, {Varadi}, {Vecchiato}, {Veljanoski}, {Via}, {Vicente}, {Vogt},
  {Voss}, {Votruba}, {Voutsinas}, {Walmsley}, {Weiler}, {Weingrill}, {Wevers},
  {Wyrzykowski}, {Yoldas}, {{\v{Z}}erjal}, {Zucker}, {Zurbach}, {Zwitter},
  {Alecu}, {Allen}, {Allende Prieto}, {Amorim}, {Anglada-Escud{\'e}},
  {Arsenijevic}, {Azaz}, {Balm}, {Beck}, {Bernstein}, {Bigot}, {Bijaoui},
  {Blasco}, {Bonfigli}, {Bono}, {Boudreault}, {Bressan}, {Brown}, {Brunet},
  {Bunclark}, {Buonanno}, {Butkevich}, {Carret}, {Carrion}, {Chemin},
  {Ch{\'e}reau}, {Corcione}, {Darmigny}, {de Boer}, {de Teodoro}, {de Zeeuw},
  {Delle Luche}, {Domingues}, {Dubath}, {Fodor}, {Fr{\'e}zouls}, {Fries},
  {Fustes}, {Fyfe}, {Gallardo}, {Gallegos}, {Gardiol}, {Gebran}, {Gomboc},
  {G{\'o}mez}, {Grux}, {Gueguen}, {Heyrovsky}, {Hoar}, {Iannicola}, {Isasi
  Parache}, {Janotto}, {Joliet}, {Jonckheere}, {Keil}, {Kim}, {Klagyivik},
  {Klar}, {Knude}, {Kochukhov}, {Kolka}, {Kos}, {Kutka}, {Lainey}, {LeBouquin},
  {Liu}, {Loreggia}, {Makarov}, {Marseille}, {Martayan}, {Martinez-Rubi},
  {Massart}, {Meynadier}, {Mignot}, {Munari}, {Nguyen}, {Nordlander}, {Ocvirk},
  {O'Flaherty}, {Olias Sanz}, {Ortiz}, {Osorio}, {Oszkiewicz}, {Ouzounis},
  {Palmer}, {Park}, {Pasquato}, {Peltzer}, {Peralta}, {P{\'e}turaud},
  {Pieniluoma}, {Pigozzi}, {Poels}, {Prat}, {Prod'homme}, {Raison}, {Rebordao},
  {Risquez}, {Rocca-Volmerange}, {Rosen}, {Ruiz-Fuertes}, {Russo}, {Sembay},
  {Serraller Vizcaino}, {Short}, {Siebert}, {Silva}, {Sinachopoulos}, {Slezak},
  {Soffel}, {Sosnowska}, {Strai{\v{z}}ys}, {ter Linden}, {Terrell}, {Theil},
  {Tiede}, {Troisi}, {Tsalmantza}, {Tur}, {Vaccari}, {Vachier}, {Valles}, {Van
  Hamme}, {Veltz}, {Virtanen}, {Wallut}, {Wichmann}, {Wilkinson}, {Ziaeepour},
  \& {Zschocke}}]{2016A&A...595A...2G}
{Gaia Collaboration}, {Brown}, A.~G.~A., {Vallenari}, A., {et~al.} 2016, \aap,
  595, A2, \dodoi{10.1051/0004-6361/201629512}

\bibitem[{{Gale} \& {Wandel}(2017)}]{2017IJAsB..16....1G}
{Gale}, J., \& {Wandel}, A. 2017, International Journal of Astrobiology, 16, 1,
  \dodoi{10.1017/S1473550415000440}

\bibitem[{Gan \& Bryant(2015)}]{gan2015adaptive}
Gan, F., \& Bryant, D.~A. 2015, Environmental microbiology, 17, 3450

\bibitem[{{Ganzeveld} \& {Lelieveld}(1995)}]{1995JGR...10020999G}
{Ganzeveld}, L., \& {Lelieveld}, J. 1995, \jgr, 100, 20,999,
  \dodoi{10.1029/95JD02266}

\bibitem[{{Gao} {et~al.}(2015){Gao}, {Hu}, {Robinson}, {Li}, \&
  {Yung}}]{2015ApJ...806..249G}
{Gao}, P., {Hu}, R., {Robinson}, T.~D., {Li}, C., \& {Yung}, Y.~L. 2015, \apj,
  806, 249, \dodoi{10.1088/0004-637X/806/2/249}

\bibitem[{{Garcia} \& {Randel}(2008)}]{2008JAtS...65.2731G}
{Garcia}, R.~R., \& {Randel}, W.~J. 2008, Journal of Atmospheric Sciences, 65,
  2731, \dodoi{10.1175/2008JAS2712.1}

\bibitem[{{Gebauer} {et~al.}(2018){Gebauer}, {Grenfell}, {Lehmann}, \&
  {Rauer}}]{2018AsBio..18..856G}
{Gebauer}, S., {Grenfell}, J.~L., {Lehmann}, R., \& {Rauer}, H. 2018,
  Astrobiology, 18, 856, \dodoi{10.1089/ast.2017.1723}

\bibitem[{{Gettelman} {et~al.}(2019){Gettelman}, {Mills}, {Kinnison}, {Garcia},
  {Smith}, {Marsh}, {Tilmes}, {Vitt}, {Bardeen}, {McInerny}, {Liu}, {Solomon},
  {Polvani}, {Emmons}, {Lamarque}, {Richter}, {Glanville}, {Bacmeister},
  {Phillips}, {Neale}, {Simpson}, {DuVivier}, {Hodzic}, \&
  {Randel}}]{2019JGRD..12412380G}
{Gettelman}, A., {Mills}, M.~J., {Kinnison}, D.~E., {et~al.} 2019, Journal of
  Geophysical Research (Atmospheres), 124, 12,380, \dodoi{10.1029/2019JD030943}

\bibitem[{{Giannakopoulos} {et~al.}(1999){Giannakopoulos}, {Chipperfield},
  {Law}, \& {Pyle}}]{1999JGR...10423761G}
{Giannakopoulos}, C., {Chipperfield}, M.~P., {Law}, K.~S., \& {Pyle}, J.~A.
  1999, \jgr, 104, 23,761, \dodoi{10.1029/1999JD900392}

\bibitem[{Giese \& Christensen(1954)}]{giese1954effects}
Giese, A.~C., \& Christensen, E. 1954, Physiological zoology, 27, 101

\bibitem[{Gon{\c{c}}alves \& Gagnon(2011)}]{gonccalves2011ozone}
Gon{\c{c}}alves, A.~A., \& Gagnon, G.~A. 2011, Ozone: Science \& Engineering,
  33, 345

\bibitem[{{Greene} {et~al.}(2023){Greene}, {Bell}, {Ducrot}, {Dyrek}, {Lagage},
  \& {Fortney}}]{2023Natur.618...39G}
{Greene}, T.~P., {Bell}, T.~J., {Ducrot}, E., {et~al.} 2023, \nat, 618, 39,
  \dodoi{10.1038/s41586-023-05951-7}

\bibitem[{{Grenfell} {et~al.}(2006){Grenfell}, {Stracke}, {Patzer}, {Titz}, \&
  {Rauer}}]{2006IJAsB...5..295G}
{Grenfell}, J.~L., {Stracke}, B., {Patzer}, B., {Titz}, R., \& {Rauer}, H.
  2006, International Journal of Astrobiology, 5, 295,
  \dodoi{10.1017/S1473550406003478}

\bibitem[{{Grenfell} {et~al.}(2012){Grenfell}, {Grie{\ss}meier}, {von Paris},
  {Patzer}, {Lammer}, {Stracke}, {Gebauer}, {Schreier}, \&
  {Rauer}}]{2012AsBio..12.1109G}
{Grenfell}, J.~L., {Grie{\ss}meier}, J.-M., {von Paris}, P., {et~al.} 2012,
  Astrobiology, 12, 1109, \dodoi{10.1089/ast.2011.0682}

\bibitem[{{Grenfell} {et~al.}(2013){Grenfell}, {Gebauer}, {Godolt},
  {Palczynski}, {Rauer}, {Stock}, {von Paris}, {Lehmann}, \&
  {Selsis}}]{2013AsBio..13..415G}
{Grenfell}, J.~L., {Gebauer}, S., {Godolt}, M., {et~al.} 2013, Astrobiology,
  13, 415, \dodoi{10.1089/ast.2012.0926}

\bibitem[{{Grimm} {et~al.}(2018){Grimm}, {Demory}, {Gillon}, {Dorn}, {Agol},
  {Burdanov}, {Delrez}, {Sestovic}, {Triaud}, {Turbet}, {Bolmont}, {Caldas},
  {de Wit}, {Jehin}, {Leconte}, {Raymond}, {Van Grootel}, {Burgasser}, {Carey},
  {Fabrycky}, {Heng}, {Hernandez}, {Ingalls}, {Lederer}, {Selsis}, \&
  {Queloz}}]{2018A&A...613A..68G}
{Grimm}, S.~L., {Demory}, B.-O., {Gillon}, M., {et~al.} 2018, \aap, 613, A68,
  \dodoi{10.1051/0004-6361/201732233}

\bibitem[{Gumsley {et~al.}(2017)Gumsley, Chamberlain, Bleeker, S{\"o}derlund,
  De~Kock, Larsson, \& Bekker}]{gumsley2017timing}
Gumsley, A.~P., Chamberlain, K.~R., Bleeker, W., {et~al.} 2017, Proceedings of
  the National Academy of Sciences, 114, 1811

\bibitem[{Guzel-Seydim {et~al.}(2004)Guzel-Seydim, Greene, \&
  Seydim}]{guzel2004use}
Guzel-Seydim, Z.~B., Greene, A.~K., \& Seydim, A. 2004, LWT-Food Science and
  Technology, 37, 453

\bibitem[{Haagen-Smit(1952)}]{haagen1952chemistry}
Haagen-Smit, A.~J. 1952, Industrial \& Engineering Chemistry, 44, 1342

\bibitem[{{Hanelt} {et~al.}(1997){Hanelt}, {Melchersmann}, {Wiencke}, \&
  {Nultsch}}]{1997MEPS..149..255H}
{Hanelt}, D., {Melchersmann}, B., {Wiencke}, C., \& {Nultsch}, W. 1997, Marine
  Ecology Progress Series, 149, 255, \dodoi{10.3354/meps149255}

\bibitem[{{Harman} {et~al.}(2018){Harman}, {Felton}, {Hu}, {Domagal-Goldman},
  {Segura}, {Tian}, \& {Kasting}}]{2018ApJ...866...56H}
{Harman}, C.~E., {Felton}, R., {Hu}, R., {et~al.} 2018, \apj, 866, 56,
  \dodoi{10.3847/1538-4357/aadd9b}

\bibitem[{{Harman} {et~al.}(2015){Harman}, {Schwieterman}, {Schottelkotte}, \&
  {Kasting}}]{2015ApJ...812..137H}
{Harman}, C.~E., {Schwieterman}, E.~W., {Schottelkotte}, J.~C., \& {Kasting},
  J.~F. 2015, \apj, 812, 137, \dodoi{10.1088/0004-637X/812/2/137}

\bibitem[{{Helmig} {et~al.}(2007){Helmig}, {Ganzeveld}, {Butler}, \&
  {Oltmans}}]{2007ACP.....7...15H}
{Helmig}, D., {Ganzeveld}, L., {Butler}, T., \& {Oltmans}, S.~J. 2007,
  Atmospheric Chemistry \& Physics, 7, 15, \dodoi{10.5194/acp-7-15-2007}

\bibitem[{Henshaw \& Laybourn-Parry(2002)}]{henshaw2002annual}
Henshaw, T., \& Laybourn-Parry, J. 2002, Polar Biology, 25, 744

\bibitem[{Hu {et~al.}(2003)Hu, Liu, \& Pei}]{hu2003characteristic}
Hu, W., Liu, P., \& Pei, H. 2003, Chinese Science Bulletin, 48, 862

\bibitem[{{Hu} \& {Yang}(2014)}]{2014PNAS..111..629H}
{Hu}, Y., \& {Yang}, J. 2014, Proceedings of the National Academy of Science,
  111, 629, \dodoi{10.1073/pnas.1315215111}

\bibitem[{Huovinen \& G{\'o}mez(2013)}]{huovinen2013photosynthetic}
Huovinen, P., \& G{\'o}mez, I. 2013, Polar biology, 36, 1319

\bibitem[{{Iriti} \& {Faoro}(2007)}]{2007WASP..187..285I}
{Iriti}, M., \& {Faoro}, F. 2007, Water Air and Soil Pollution, 187, 285,
  \dodoi{10.1007/s11270-007-9517-7}

\bibitem[{{Ji} {et~al.}(2023){Ji}, {Kasting}, {Cooke}, {Marsh}, \&
  {Tsigaridis}}]{2023RSOS...1030056J}
{Ji}, A., {Kasting}, J.~F., {Cooke}, G.~J., {Marsh}, D.~R., \& {Tsigaridis}, K.
  2023, Royal Society Open Science, 10, 230056, \dodoi{10.1098/rsos.230056}

\bibitem[{Ji {et~al.}(2024)Ji, Tomazzeli, Palancar, Chaverot, Barker,
  Fern{\'a}ndez, Minschwaner, \& Kasting}]{ji2024correlated}
Ji, A., Tomazzeli, O.~G., Palancar, G.~G., {et~al.} 2024, Journal of
  Geophysical Research: Atmospheres, 129, e2023JD040610,
  \dodoi{10.1029/2023JD040610}

\bibitem[{Jones {et~al.}(2006)Jones, Gensemer, Stubblefield, Van~Genderen,
  Dethloff, \& Cooper}]{jones2006toxicity}
Jones, A.~C., Gensemer, R.~W., Stubblefield, W.~A., {et~al.} 2006,
  Environmental Toxicology and Chemistry: An International Journal, 25, 2683

\bibitem[{{Joshi} {et~al.}(1997){Joshi}, {Haberle}, \&
  {Reynolds}}]{1997Icar..129..450J}
{Joshi}, M.~M., {Haberle}, R.~M., \& {Reynolds}, R.~T. 1997, \icarus, 129, 450,
  \dodoi{10.1006/icar.1997.5793}

\bibitem[{{Jung} {et~al.}(2023){Jung}, {Harion}, {Wu}, {N{\"u}rnberg},
  {Bellamoli}, {Guillen}, {Leira}, \& {Lakatos}}]{2023FrASS..1007371J}
{Jung}, P., {Harion}, F., {Wu}, S., {et~al.} 2023, Frontiers in Astronomy and
  Space Sciences, 10, 5, \dodoi{10.3389/fspas.2023.1107371}

\bibitem[{{Kaltenegger} {et~al.}(2020){Kaltenegger}, {Lin}, \&
  {Rugheimer}}]{2020ApJ...904...10K}
{Kaltenegger}, L., {Lin}, Z., \& {Rugheimer}, S. 2020, \apj, 904, 10,
  \dodoi{10.3847/1538-4357/abb9b2}

\bibitem[{{Kaltenegger} \& {Sasselov}(2010)}]{2010ApJ...708.1162K}
{Kaltenegger}, L., \& {Sasselov}, D. 2010, \apj, 708, 1162,
  \dodoi{10.1088/0004-637X/708/2/1162}

\bibitem[{Kappen(1993)}]{kappen1993plant}
Kappen, L. 1993, Arctic, 297

\bibitem[{{Kasting} {et~al.}(1993){Kasting}, {Whitmire}, \&
  {Reynolds}}]{1993Icar..101..108K}
{Kasting}, J.~F., {Whitmire}, D.~P., \& {Reynolds}, R.~T. 1993, \icarus, 101,
  108, \dodoi{10.1006/icar.1993.1010}

\bibitem[{Kim {et~al.}(1999)Kim, Yousef, \& Dave}]{kim1999application}
Kim, J.-G., Yousef, A.~E., \& Dave, S. 1999, Journal of food protection, 62,
  1071

\bibitem[{Kishimoto \& Arai(2022)}]{kishimoto2022effect}
Kishimoto, N., \& Arai, H. 2022, Ozone: Science \& Engineering, 44, 265

\bibitem[{Klaunig {et~al.}(2010)Klaunig, Kamendulis, \&
  Hocevar}]{klaunig2010oxidative}
Klaunig, J.~E., Kamendulis, L.~M., \& Hocevar, B.~A. 2010, Toxicologic
  pathology, 38, 96

\bibitem[{{Kleinb{\"o}hl} {et~al.}(2018){Kleinb{\"o}hl}, {Willacy}, {Friedson},
  {Chen}, \& {Swain}}]{2018ApJ...862...92K}
{Kleinb{\"o}hl}, A., {Willacy}, K., {Friedson}, A.~J., {Chen}, P., \& {Swain},
  M.~R. 2018, \apj, 862, 92, \dodoi{10.3847/1538-4357/aaca36}

\bibitem[{{Kopparapu} {et~al.}(2021){Kopparapu}, {Arney}, {Haqq-Misra},
  {Lustig-Yaeger}, \& {Villanueva}}]{2021ApJ...908..164K}
{Kopparapu}, R., {Arney}, G., {Haqq-Misra}, J., {Lustig-Yaeger}, J., \&
  {Villanueva}, G. 2021, \apj, 908, 164, \dodoi{10.3847/1538-4357/abd7f7}

\bibitem[{{Kozakis} \& {Kaltenegger}(2020)}]{2020AJ....160..225K}
{Kozakis}, T., \& {Kaltenegger}, L. 2020, \aj, 160, 225,
  \dodoi{10.3847/1538-3881/abb9ac}

\bibitem[{{Kozakis} {et~al.}(2018){Kozakis}, {Kaltenegger}, \&
  {Hoard}}]{2018ApJ...862...69K}
{Kozakis}, T., {Kaltenegger}, L., \& {Hoard}, D.~W. 2018, \apj, 862, 69,
  \dodoi{10.3847/1538-4357/aacbc7}

\bibitem[{{Kozakis} {et~al.}(2022){Kozakis}, {Mendon{\c{c}}a}, \&
  {Buchhave}}]{2022A&A...665A.156K}
{Kozakis}, T., {Mendon{\c{c}}a}, J.~M., \& {Buchhave}, L.~A. 2022, \aap, 665,
  A156, \dodoi{10.1051/0004-6361/202244164}

\bibitem[{{Large} {et~al.}(2019){Large}, {Mukherjee}, {Gregory}, {Steadman},
  {Corkrey}, \& {Danyushevsky}}]{2019MinDe..54..485L}
{Large}, R.~R., {Mukherjee}, I., {Gregory}, D., {et~al.} 2019, Mineralium
  Deposita, 54, 485, \dodoi{10.1007/s00126-019-00873-9}

\bibitem[{{Leger} {et~al.}(1993){Leger}, {Pirre}, \&
  {Marceau}}]{1993A&A...277..309L}
{Leger}, A., {Pirre}, M., \& {Marceau}, F.~J. 1993, \aap, 277, 309

\bibitem[{{Lehmer} {et~al.}(2018){Lehmer}, {Catling}, {Parenteau}, \&
  {Hoehler}}]{2018ApJ...859..171L}
{Lehmer}, O.~R., {Catling}, D.~C., {Parenteau}, M.~N., \& {Hoehler}, T.~M.
  2018, \apj, 859, 171, \dodoi{10.3847/1538-4357/aac104}

\bibitem[{{Lewis} {et~al.}(2018){Lewis}, {Lambert}, {Boutle}, {Mayne},
  {Manners}, \& {Acreman}}]{2018ApJ...854..171L}
{Lewis}, N.~T., {Lambert}, F.~H., {Boutle}, I.~A., {et~al.} 2018, \apj, 854,
  171, \dodoi{10.3847/1538-4357/aaad0a}

\bibitem[{{Lin} {et~al.}(2021){Lin}, {MacDonald}, {Kaltenegger}, \&
  {Wilson}}]{2021MNRAS.505.3562L}
{Lin}, Z., {MacDonald}, R.~J., {Kaltenegger}, L., \& {Wilson}, D.~J. 2021,
  \mnras, 505, 3562, \dodoi{10.1093/mnras/stab1486}

\bibitem[{{Lingam} \& {Loeb}(2019)}]{2019MNRAS.485.5924L}
{Lingam}, M., \& {Loeb}, A. 2019, \mnras, 485, 5924,
  \dodoi{10.1093/mnras/stz847}

\bibitem[{Loose {et~al.}(2011)Loose, Miller, Elliott, \&
  Papakyriakou}]{loose2011sea}
Loose, B., Miller, L.~A., Elliott, S., \& Papakyriakou, T. 2011, Oceanography,
  24, 202

\bibitem[{{Loose} {et~al.}(2011){Loose}, {Schlosser}, {Perovich}, {Ringelberg},
  {Ho}, {Takahashi}, {Richter-Menge}, {Reynolds}, {Mcgillis}, \&
  {Tison}}]{2011TellB..63...23L}
{Loose}, B., {Schlosser}, P., {Perovich}, D., {et~al.} 2011, Tellus Series B
  Chemical and Physical Meteorology B, 63, 23,
  \dodoi{10.1111/j.1600-0889.2010.00506.x}

\bibitem[{{Loyd} {et~al.}(2016){Loyd}, {France}, {Youngblood}, {Schneider},
  {Brown}, {Hu}, {Linsky}, {Froning}, {Redfield}, {Rugheimer}, \&
  {Tian}}]{2016ApJ...824..102L}
{Loyd}, R.~O.~P., {France}, K., {Youngblood}, A., {et~al.} 2016, \apj, 824,
  102, \dodoi{10.3847/0004-637X/824/2/102}

\bibitem[{{Luger} \& {Barnes}(2015)}]{2015AAS...22540704L}
{Luger}, R., \& {Barnes}, R. 2015, in American Astronomical Society Meeting
  Abstracts, Vol. 225, American Astronomical Society Meeting Abstracts \#225,
  407.04

\bibitem[{Lushchak(2011)}]{lushchak2011environmentally}
Lushchak, V.~I. 2011, Aquatic toxicology, 101, 13

\bibitem[{Lykkesfeldt \& Svendsen(2007)}]{lykkesfeldt2007oxidants}
Lykkesfeldt, J., \& Svendsen, O. 2007, The veterinary journal, 173, 502

\bibitem[{{Lyons} {et~al.}(2021){Lyons}, {Diamond}, {Planavsky}, {Reinhard}, \&
  {Li}}]{2021AsBio..21..906L}
{Lyons}, T.~W., {Diamond}, C.~W., {Planavsky}, N.~J., {Reinhard}, C.~T., \&
  {Li}, C. 2021, Astrobiology, 21, 906, \dodoi{10.1089/ast.2020.2418}

\bibitem[{{Lyons} {et~al.}(2014){Lyons}, {Reinhard}, \&
  {Planavsky}}]{2014Natur.506..307L}
{Lyons}, T.~W., {Reinhard}, C.~T., \& {Planavsky}, N.~J. 2014, \nat, 506, 307,
  \dodoi{10.1038/nature13068}

\bibitem[{{Macdonald} {et~al.}(2022){Macdonald}, {Paradise}, {Menou}, \&
  {Lee}}]{2022MNRAS.513.2761M}
{Macdonald}, E., {Paradise}, A., {Menou}, K., \& {Lee}, C. 2022, \mnras, 513,
  2761, \dodoi{10.1093/mnras/stac1040}

\bibitem[{{Madhusudhan} {et~al.}(2021){Madhusudhan}, {Piette}, \&
  {Constantinou}}]{2021ApJ...918....1M}
{Madhusudhan}, N., {Piette}, A. A.~A., \& {Constantinou}, S. 2021, \apj, 918,
  1, \dodoi{10.3847/1538-4357/abfd9c}

\bibitem[{Malashock {et~al.}(2022)Malashock, Delang, Becker, Serre, West,
  Chang, Cooper, \& Anenberg}]{malashock2022global}
Malashock, D.~A., Delang, M.~N., Becker, J.~S., {et~al.} 2022, The Lancet
  Planetary Health, 6, e958

\bibitem[{{Malashock} {et~al.}(2022){Malashock}, {DeLang}, {Becker}, {Serre},
  {West}, {Chang}, {Cooper}, \& {Anenberg}}]{2022ERL....17e4023M}
{Malashock}, D.~A., {DeLang}, M.~N., {Becker}, J.~S., {et~al.} 2022,
  Environmental Research Letters, 17, 054023, \dodoi{10.1088/1748-9326/ac66f3}

\bibitem[{{Marcos-Arenal} {et~al.}(2022){Marcos-Arenal}, {Cerd{\'a}n},
  {Burillo-Villalobos}, {Fonseca-Bonilla}, {de la Concepci{\'o}n},
  {L{\'o}pez-Cayuela}, {G{\'o}mez}, \& {Caballero}}]{2022Univ....8..624M}
{Marcos-Arenal}, P., {Cerd{\'a}n}, L., {Burillo-Villalobos}, M., {et~al.} 2022,
  Universe, 8, 624, \dodoi{10.3390/universe8120624}

\bibitem[{McMinn {et~al.}(2007)McMinn, Ryan, Ralph, \&
  Pankowski}]{mcminn2007spring}
McMinn, A., Ryan, K., Ralph, P., \& Pankowski, A. 2007, Marine Biology, 151,
  985

\bibitem[{McNaught {et~al.}(1997)McNaught, Wilkinson,
  {et~al.}}]{mcnaught1997compendium}
McNaught, A.~D., Wilkinson, A., {et~al.} 1997, Compendium of chemical
  terminology, Vol. 1669 (Blackwell Science Oxford)

\bibitem[{{Meadows} {et~al.}(2018){Meadows}, {Arney}, {Schwieterman},
  {Lustig-Yaeger}, {Lincowski}, {Robinson}, {Domagal-Goldman}, {Deitrick},
  {Barnes}, {Fleming}, {Luger}, {Driscoll}, {Quinn}, \&
  {Crisp}}]{2018AsBio..18..133M}
{Meadows}, V.~S., {Arney}, G.~N., {Schwieterman}, E.~W., {et~al.} 2018,
  Astrobiology, 18, 133, \dodoi{10.1089/ast.2016.1589}

\bibitem[{{M{\'e}ndez} {et~al.}(2021){M{\'e}ndez}, {Rivera-Valent{\'\i}n},
  {Schulze-Makuch}, {Filiberto}, {Ram{\'\i}rez}, {Wood}, {D{\'a}vila}, {McKay},
  {Ceballos}, {Jusino-Maldonado}, {Torres-Santiago}, {Nery}, {Heller}, {Byrne},
  {Malaska}, {Nathan}, {Sim{\~o}es}, {Antunes}, {Mart{\'\i}nez-Fr{\'\i}as},
  {Carone}, {Izenberg}, {Atri}, {Chitty}, {Nowajewski-Barra},
  {Rivera-Hern{\'a}ndez}, {Brown}, {Lynch}, {Catling}, {Zuluaga}, {Salazar},
  {Chen}, {Gonz{\'a}lez}, {Jagadeesh}, \& {Haqq-Misra}}]{2021AsBio..21.1017M}
{M{\'e}ndez}, A., {Rivera-Valent{\'\i}n}, E.~G., {Schulze-Makuch}, D., {et~al.}
  2021, Astrobiology, 21, 1017, \dodoi{10.1089/ast.2020.2342}

\bibitem[{Menzel(1984)}]{menzel1984ozone}
Menzel, D.~B. 1984, Journal of Toxicology and Environmental Health

\bibitem[{Najafi \& Khodaparast(2009)}]{najafi2009efficacy}
Najafi, M. B.~H., \& Khodaparast, M.~H. 2009, Food control, 20, 27

\bibitem[{Norris \& Olsen(1991)}]{norris1991deep}
Norris, J., \& Olsen, J. 1991, Phycologia, 30, 315

\bibitem[{Nuvolone {et~al.}(2018)Nuvolone, Petri, \&
  Voller}]{nuvolone2018effects}
Nuvolone, D., Petri, D., \& Voller, F. 2018, Environmental Science and
  Pollution Research, 25, 8074

\bibitem[{{Olson} {et~al.}(2022){Olson}, {Jansen}, {Abbot}, {Halevy}, \&
  {Goldblatt}}]{2022GeoRL..4995748O}
{Olson}, S., {Jansen}, M.~F., {Abbot}, D.~S., {Halevy}, I., \& {Goldblatt}, C.
  2022, \grl, 49, e95748, \dodoi{10.1029/2021GL095748}

\bibitem[{{Otegi} {et~al.}(2020){Otegi}, {Bouchy}, \&
  {Helled}}]{2020A&A...634A..43O}
{Otegi}, J.~F., {Bouchy}, F., \& {Helled}, R. 2020, \aap, 634, A43,
  \dodoi{10.1051/0004-6361/201936482}

\bibitem[{{Peacock}(2020)}]{t9-j6bz-5g89}
{Peacock}, S. 2020, {Habitable Zones and M dwarf Activity across Time
  ("HAZMAT")}, Version 1,  STScI/MAST, \dodoi{10.17909/T9-J6BZ-5G89}

\bibitem[{{Peacock} {et~al.}(2019){Peacock}, {Barman}, {Shkolnik},
  {Hauschildt}, \& {Baron}}]{2019ApJ...871..235P}
{Peacock}, S., {Barman}, T., {Shkolnik}, E.~L., {Hauschildt}, P.~H., \&
  {Baron}, E. 2019, \apj, 871, 235, \dodoi{10.3847/1538-4357/aaf891}

\bibitem[{{Pidhorodetska} {et~al.}(2020){Pidhorodetska}, {Fauchez},
  {Villanueva}, {Domagal-Goldman}, \& {Kopparapu}}]{2020ApJ...898L..33P}
{Pidhorodetska}, D., {Fauchez}, T.~J., {Villanueva}, G.~L., {Domagal-Goldman},
  S.~D., \& {Kopparapu}, R.~K. 2020, \apjl, 898, L33,
  \dodoi{10.3847/2041-8213/aba4a1}

\bibitem[{{Pierrehumbert} \& {Hammond}(2019)}]{2019AnRFM..51..275P}
{Pierrehumbert}, R.~T., \& {Hammond}, M. 2019, Annual Review of Fluid
  Mechanics, 51, 275, \dodoi{10.1146/annurev-fluid-010518-040516}

\bibitem[{{Pineda} {et~al.}(2021){Pineda}, {Youngblood}, \&
  {France}}]{2021ApJ...918...40P}
{Pineda}, J.~S., {Youngblood}, A., \& {France}, K. 2021, \apj, 918, 40,
  \dodoi{10.3847/1538-4357/ac0aea}

\bibitem[{Poulton {et~al.}(2021)Poulton, Bekker, Cumming, Zerkle, Canfield, \&
  Johnston}]{poulton2021200}
Poulton, S.~W., Bekker, A., Cumming, V.~M., {et~al.} 2021, Nature, 592, 232

\bibitem[{Premjit {et~al.}(2022)Premjit, Sruthi, Pandiselvam, \&
  Kothakota}]{premjit2022aqueous}
Premjit, Y., Sruthi, N., Pandiselvam, R., \& Kothakota, A. 2022, Comprehensive
  Reviews in Food Science and Food Safety, 21, 1054

\bibitem[{Pritchard {et~al.}(2013)Pritchard, Hurd, Beardall, \&
  Hepburn}]{pritchard2013survival}
Pritchard, D.~W., Hurd, C.~L., Beardall, J., \& Hepburn, C.~D. 2013, Journal of
  Phycology, 49, 867

\bibitem[{{Proedrou} \& {Hocke}(2016)}]{2016EP&S...68...96P}
{Proedrou}, E., \& {Hocke}, K. 2016, Earth, Planets and Space, 68, 96,
  \dodoi{10.1186/s40623-016-0461-x}

\bibitem[{{Rajpurohit} {et~al.}(2013){Rajpurohit}, {Reyl{\'e}}, {Allard},
  {Homeier}, {Schultheis}, {Bessell}, \& {Robin}}]{2013A&A...556A..15R}
{Rajpurohit}, A.~S., {Reyl{\'e}}, C., {Allard}, F., {et~al.} 2013, \aap, 556,
  A15, \dodoi{10.1051/0004-6361/201321346}

\bibitem[{{Ramirez}(2020)}]{2020NatSR..10.7432R}
{Ramirez}, R.~M. 2020, Scientific Reports, 10, 7432,
  \dodoi{10.1038/s41598-020-64436-z}

\bibitem[{Ramus {et~al.}(1976)Ramus, Beale, \&
  Mauzerall}]{ramus1976correlation}
Ramus, J., Beale, S., \& Mauzerall, D. 1976, Marine Biology, 37, 231

\bibitem[{{Ramya} {et~al.}(2023){Ramya}, {Dhevagi}, {Poornima},
  {Avudainayagam}, {Watanabe}, \& {Agathokleous}}]{2023ER....236k6816R}
{Ramya}, A., {Dhevagi}, P., {Poornima}, R., {et~al.} 2023, Environmental
  Research, 236, 116816, \dodoi{10.1016/j.envres.2023.116816}

\bibitem[{Rao \& Davis(2001)}]{rao2001physiology}
Rao, M.~V., \& Davis, K.~R. 2001, Planta, 213, 682

\bibitem[{{Reinhard} {et~al.}(2017){Reinhard}, {Olson}, {Schwieterman}, \&
  {Lyons}}]{2017AsBio..17..287R}
{Reinhard}, C.~T., {Olson}, S.~L., {Schwieterman}, E.~W., \& {Lyons}, T.~W.
  2017, Astrobiology, 17, 287, \dodoi{10.1089/ast.2016.1598}

\bibitem[{{Renaud} {et~al.}(2021){Renaud}, {Henning}, {Saxena}, {Neveu},
  {Bagheri}, {Mandell}, \& {Hurford}}]{2021PSJ.....2....4R}
{Renaud}, J.~P., {Henning}, W.~G., {Saxena}, P., {et~al.} 2021, \psj, 2, 4,
  \dodoi{10.3847/PSJ/abc0f3}

\bibitem[{{Ribas} {et~al.}(2016){Ribas}, {Bolmont}, {Selsis}, {Reiners},
  {Leconte}, {Raymond}, {Engle}, {Guinan}, {Morin}, {Turbet}, {Forget}, \&
  {Anglada-Escud{\'e}}}]{2016A&A...596A.111R}
{Ribas}, I., {Bolmont}, E., {Selsis}, F., {et~al.} 2016, \aap, 596, A111,
  \dodoi{10.1051/0004-6361/201629576}

\bibitem[{Rich(1964)}]{rich1964ozone}
Rich, S. 1964, Annual Review of Phytopathology, 2, 253

\bibitem[{{Ridgway} {et~al.}(2023){Ridgway}, {Zamyatina}, {Mayne}, {Manners},
  {Lambert}, {Braam}, {Drummond}, {H{\'e}brard}, {Palmer}, \&
  {Kohary}}]{2023MNRAS.518.2472R}
{Ridgway}, R.~J., {Zamyatina}, M., {Mayne}, N.~J., {et~al.} 2023, \mnras, 518,
  2472, \dodoi{10.1093/mnras/stac3105}

\bibitem[{Rojas-Valencia(2011)}]{rojas2011research}
Rojas-Valencia, M. 2011, Virus, 3, 263

\bibitem[{{Rugheimer} \& {Kaltenegger}(2018)}]{2018ApJ...854...19R}
{Rugheimer}, S., \& {Kaltenegger}, L. 2018, \apj, 854, 19,
  \dodoi{10.3847/1538-4357/aaa47a}

\bibitem[{{Rugheimer} {et~al.}(2015){Rugheimer}, {Kaltenegger}, {Segura},
  {Linsky}, \& {Mohanty}}]{2015ApJ...809...57R}
{Rugheimer}, S., {Kaltenegger}, L., {Segura}, A., {Linsky}, J., \& {Mohanty},
  S. 2015, \apj, 809, 57, \dodoi{10.1088/0004-637X/809/1/57}

\bibitem[{{Safonova} {et~al.}(2021){Safonova}, {Mathur}, {Basak}, {Bora}, \&
  {Agrawal}}]{2021EPJST.230.2207S}
{Safonova}, M., {Mathur}, A., {Basak}, S., {Bora}, K., \& {Agrawal}, S. 2021,
  European Physical Journal Special Topics, 230, 2207,
  \dodoi{10.1140/epjs/s11734-021-00211-z}

\bibitem[{{Salazar} {et~al.}(2020){Salazar}, {Olson}, {Komacek}, {Stephens}, \&
  {Abbot}}]{2020ApJ...896L..16S}
{Salazar}, A.~M., {Olson}, S.~L., {Komacek}, T.~D., {Stephens}, H., \& {Abbot},
  D.~S. 2020, \apjl, 896, L16, \dodoi{10.3847/2041-8213/ab94c1}

\bibitem[{Sandermann~Jr(1996)}]{sandermann1996ozone}
Sandermann~Jr, H. 1996, Annual review of phytopathology, 34, 347

\bibitem[{Sandu {et~al.}(1997)Sandu, Verwer, Blom, Spee, Carmichael, \&
  Potra}]{sandu1997benchmarking}
Sandu, A., Verwer, J., Blom, J., {et~al.} 1997, Atmospheric environment, 31,
  3459

\bibitem[{{Scheucher} {et~al.}(2020){Scheucher}, {Herbst}, {Schmidt},
  {Grenfell}, {Schreier}, {Banjac}, {Heber}, {Rauer}, \&
  {Sinnhuber}}]{2020ApJ...893...12S}
{Scheucher}, M., {Herbst}, K., {Schmidt}, V., {et~al.} 2020, \apj, 893, 12,
  \dodoi{10.3847/1538-4357/ab7b74}

\bibitem[{{Schwieterman} {et~al.}(2019){Schwieterman}, {Reinhard}, {Olson},
  {Harman}, \& {Lyons}}]{2019ApJ...878...19S}
{Schwieterman}, E.~W., {Reinhard}, C.~T., {Olson}, S.~L., {Harman}, C.~E., \&
  {Lyons}, T.~W. 2019, \apj, 878, 19, \dodoi{10.3847/1538-4357/ab1d52}

\bibitem[{{Schwieterman} {et~al.}(2016){Schwieterman}, {Meadows},
  {Domagal-Goldman}, {Deming}, {Arney}, {Luger}, {Harman}, {Misra}, \&
  {Barnes}}]{2016ApJ...819L..13S}
{Schwieterman}, E.~W., {Meadows}, V.~S., {Domagal-Goldman}, S.~D., {et~al.}
  2016, \apjl, 819, L13, \dodoi{10.3847/2041-8205/819/1/L13}

\bibitem[{{Schwieterman} {et~al.}(2018){Schwieterman}, {Kiang}, {Parenteau},
  {Harman}, {DasSarma}, {Fisher}, {Arney}, {Hartnett}, {Reinhard}, {Olson},
  {Meadows}, {Cockell}, {Walker}, {Grenfell}, {Hegde}, {Rugheimer}, {Hu}, \&
  {Lyons}}]{2018AsBio..18..663S}
{Schwieterman}, E.~W., {Kiang}, N.~Y., {Parenteau}, M.~N., {et~al.} 2018,
  Astrobiology, 18, 663, \dodoi{10.1089/ast.2017.1729}

\bibitem[{{Segura} {et~al.}(2007){Segura}, {Meadows}, {Kasting}, {Crisp}, \&
  {Cohen}}]{2007A&A...472..665S}
{Segura}, A., {Meadows}, V.~S., {Kasting}, J.~F., {Crisp}, D., \& {Cohen}, M.
  2007, \aap, 472, 665, \dodoi{10.1051/0004-6361:20066663}

\bibitem[{{Segura} {et~al.}(2010){Segura}, {Walkowicz}, {Meadows}, {Kasting},
  \& {Hawley}}]{2010AsBio..10..751S}
{Segura}, A., {Walkowicz}, L.~M., {Meadows}, V., {Kasting}, J., \& {Hawley}, S.
  2010, Astrobiology, 10, 751, \dodoi{10.1089/ast.2009.0376}

\bibitem[{{Sergeev} {et~al.}(2023){Sergeev}, {Mayne}, {Bendall}, {Boutle},
  {Brown}, {Kav{\v{c}}i{\v{c}}}, {Kent}, {Kohary}, {Manners}, {Melvin},
  {Olivier}, {Ragta}, {Shipway}, {Wakelin}, {Wood}, \&
  {Zerroukat}}]{2023GMD....16.5601S}
{Sergeev}, D.~E., {Mayne}, N.~J., {Bendall}, T., {et~al.} 2023, Geoscientific
  Model Development, 16, 5601, \dodoi{10.5194/gmd-16-5601-2023}

\bibitem[{Sharkey {et~al.}(2008)Sharkey, Wiberley, \&
  Donohue}]{sharkey2008isoprene}
Sharkey, T.~D., Wiberley, A.~E., \& Donohue, A.~R. 2008, Annals of botany, 101,
  5

\bibitem[{{Showman} \& {Polvani}(2011)}]{2011ApJ...738...71S}
{Showman}, A.~P., \& {Polvani}, L.~M. 2011, \apj, 738, 71,
  \dodoi{10.1088/0004-637X/738/1/71}

\bibitem[{Sies {et~al.}(2017)Sies, Berndt, \& Jones}]{sies2017oxidative}
Sies, H., Berndt, C., \& Jones, D.~P. 2017, Annual review of biochemistry, 86,
  715

\bibitem[{Sillman(1999)}]{sillman1999relation}
Sillman, S. 1999, Atmospheric Environment, 33, 1821

\bibitem[{Silva {et~al.}(2013)Silva, West, Zhang, Anenberg, Lamarque, Shindell,
  Collins, Dalsoren, Faluvegi, Folberth, {et~al.}}]{silva2013global}
Silva, R.~A., West, J.~J., Zhang, Y., {et~al.} 2013, Environmental Research
  Letters, 8, 034005

\bibitem[{{Squire} {et~al.}(2014){Squire}, {Archibald}, {Abraham}, {Beerling},
  {Hewitt}, {Lathi{\`e}re}, {Pike}, {Telford}, \& {Pyle}}]{2014ACP....14.1011S}
{Squire}, O.~J., {Archibald}, A.~T., {Abraham}, N.~L., {et~al.} 2014,
  Atmospheric Chemistry \& Physics, 14, 1011, \dodoi{10.5194/acp-14-1011-2014}

\bibitem[{{Steadman} {et~al.}(2020){Steadman}, {Large}, {Blamey}, {Mukherjee},
  {Corkrey}, {Danyushevsky}, {Maslennikov}, {Hollings}, {Garven}, {Brand}, \&
  {L{\'e}cuyer}}]{2020PreR..343j5722S}
{Steadman}, J.~A., {Large}, R.~R., {Blamey}, N.~J., {et~al.} 2020, Precambrian
  Research, 343, 105722, \dodoi{10.1016/j.precamres.2020.105722}

\bibitem[{Stokinger(1965)}]{stokinger1965ozone}
Stokinger, H. 1965, Archives of Environmental Health: An International Journal,
  10, 719

\bibitem[{{Sun} {et~al.}(2022){Sun}, {Yu}, {Lan}, {Wan}, {Hickman},
  {Murulitharan}, {Shen}, {Yuan}, {Guo}, \& {Archibald}}]{2022Innov...300246S}
{Sun}, H.~Z., {Yu}, P., {Lan}, C., {et~al.} 2022, The Innovation, 3, 100246,
  \dodoi{10.1016/j.xinn.2022.100246}

\bibitem[{{Suzuki} {et~al.}(1997){Suzuki}, {Kudoh}, \&
  {Takahashi}}]{1997JMS....11..111S}
{Suzuki}, Y., {Kudoh}, S., \& {Takahashi}, M. 1997, Journal of Marine Systems,
  11, 111, \dodoi{10.1016/S0924-7963(96)00032-2}

\bibitem[{{Tabataba-Vakili} {et~al.}(2016){Tabataba-Vakili}, {Grenfell},
  {Grie{\ss}meier}, \& {Rauer}}]{2016A&A...585A..96T}
{Tabataba-Vakili}, F., {Grenfell}, J.~L., {Grie{\ss}meier}, J.~M., \& {Rauer},
  H. 2016, \aap, 585, A96, \dodoi{10.1051/0004-6361/201425602}

\bibitem[{{Tian} {et~al.}(2014){Tian}, {France}, {Linsky}, {Mauas}, \&
  {Vieytes}}]{2014E&PSL.385...22T}
{Tian}, F., {France}, K., {Linsky}, J.~L., {Mauas}, P. J.~D., \& {Vieytes},
  M.~C. 2014, Earth and Planetary Science Letters, 385, 22,
  \dodoi{10.1016/j.epsl.2013.10.024}

\bibitem[{{Tilley} {et~al.}(2019){Tilley}, {Segura}, {Meadows}, {Hawley}, \&
  {Davenport}}]{2019AsBio..19...64T}
{Tilley}, M.~A., {Segura}, A., {Meadows}, V., {Hawley}, S., \& {Davenport}, J.
  2019, Astrobiology, 19, 64, \dodoi{10.1089/ast.2017.1794}

\bibitem[{{Tjoa} {et~al.}(2020){Tjoa}, {Mueller}, \& {van der
  Tak}}]{2020A&A...636A..50T}
{Tjoa}, J.~N.~K.~Y., {Mueller}, M., \& {van der Tak}, F.~F.~S. 2020, \aap, 636,
  A50, \dodoi{10.1051/0004-6361/201937035}

\bibitem[{{Truitt} {et~al.}(2020){Truitt}, {Young}, {Walker}, \&
  {Spacek}}]{2020AJ....159...55T}
{Truitt}, A.~R., {Young}, P.~A., {Walker}, S.~I., \& {Spacek}, A. 2020, \aj,
  159, 55, \dodoi{10.3847/1538-3881/ab4e93}

\bibitem[{{Turbet} {et~al.}(2022){Turbet}, {Fauchez}, {Sergeev}, {Boutle},
  {Tsigaridis}, {Way}, {Wolf}, {Domagal-Goldman}, {Forget}, {Haqq-Misra},
  {Kopparapu}, {Lambert}, {Manners}, {Mayne}, \& {Sohl}}]{2022PSJ.....3..211T}
{Turbet}, M., {Fauchez}, T.~J., {Sergeev}, D.~E., {et~al.} 2022, \psj, 3, 211,
  \dodoi{10.3847/PSJ/ac6cf0}

\bibitem[{Turner {et~al.}(2016)Turner, Jerrett, Pope~III, Krewski, Gapstur,
  Diver, Beckerman, Marshall, Su, Crouse, {et~al.}}]{turner2016long}
Turner, M.~C., Jerrett, M., Pope~III, C.~A., {et~al.} 2016, American journal of
  respiratory and critical care medicine, 193, 1134

\bibitem[{{Val Martin} {et~al.}(2014){Val Martin}, {Heald}, \&
  {Arnold}}]{2014GeoRL..41.2988V}
{Val Martin}, M., {Heald}, C.~L., \& {Arnold}, S.~R. 2014, \grl, 41, 2988,
  \dodoi{10.1002/2014GL059651}

\bibitem[{Valavanidis {et~al.}(2013)Valavanidis, Vlachogianni, Fiotakis, \&
  Loridas}]{valavanidis2013pulmonary}
Valavanidis, A., Vlachogianni, T., Fiotakis, K., \& Loridas, S. 2013,
  International journal of environmental research and public health, 10, 3886

\bibitem[{{Way} {et~al.}(2017){Way}, {Aleinov}, {Amundsen}, {Chandler},
  {Clune}, {Del Genio}, {Fujii}, {Kelley}, {Kiang}, {Sohl}, \&
  {Tsigaridis}}]{2017ApJS..231...12W}
{Way}, M.~J., {Aleinov}, I., {Amundsen}, D.~S., {et~al.} 2017, \apjs, 231, 12,
  \dodoi{10.3847/1538-4365/aa7a06}

\bibitem[{Wert {et~al.}(2017)Wert, Lew, \& Rakness}]{wert2017effect}
Wert, E.~C., Lew, J., \& Rakness, K.~L. 2017, Journal-American Water Works
  Association, 109, E302

\bibitem[{{Wesely}(1989)}]{1989AtmEn..23.1293W}
{Wesely}, M.~L. 1989, Atmospheric Environment, 23, 1293,
  \dodoi{10.1016/0004-6981(89)90153-4}

\bibitem[{{Wesely} \& {Hicks}(2000)}]{2000AtmEn..34.2261W}
{Wesely}, M.~L., \& {Hicks}, B.~B. 2000, Atmospheric Environment, 34, 2261,
  \dodoi{10.1016/S1352-2310(99)00467-7}

\bibitem[{{Wilson} {et~al.}(2021{\natexlab{a}}){Wilson}, {Froning}, {Duvvuri},
  {France}, {Youngblood}, \& {Schneider}}]{zenodo.4556130}
{Wilson}, D.~J., {Froning}, C., {Duvvuri}, G., {et~al.} 2021{\natexlab{a}},
  {Mega-MUSCLES Semi-empirical SED of TRAPPIST-1}, Version 1,  Zenodo.
\newblock \url{https://doi.org/10.5281/zenodo.4556130}

\bibitem[{{Wilson} {et~al.}(2021{\natexlab{b}}){Wilson}, {Froning}, {Duvvuri},
  {France}, {Youngblood}, {Schneider}, {Berta-Thompson}, {Brown}, {Buccino},
  {Hawley}, {Irwin}, {Kaltenegger}, {Kowalski}, {Linsky}, {Parke Loyd},
  {Miguel}, {Pineda}, {Redfield}, {Roberge}, {Rugheimer}, {Tian}, \&
  {Vieytes}}]{2021ApJ...911...18W}
{Wilson}, D.~J., {Froning}, C.~S., {Duvvuri}, G.~M., {et~al.}
  2021{\natexlab{b}}, \apj, 911, 18, \dodoi{10.3847/1538-4357/abe771}

\bibitem[{{Wordsworth} \& {Pierrehumbert}(2014)}]{2014ApJ...785L..20W}
{Wordsworth}, R., \& {Pierrehumbert}, R. 2014, \apjl, 785, L20,
  \dodoi{10.1088/2041-8205/785/2/L20}

\bibitem[{{World Health Organization} {et~al.}(2000)}]{world2000air}
{World Health Organization}, {et~al.} 2000, Air quality guidelines for Europe
  (World Health Organization. Regional Office for Europe)

\bibitem[{{Yassin Jaziri} {et~al.}(2022){Yassin Jaziri}, {Charnay}, {Selsis},
  {Leconte}, \& {Lef{\`e}vre}}]{2022CliPa..18.2421Y}
{Yassin Jaziri}, A., {Charnay}, B., {Selsis}, F., {Leconte}, J., \&
  {Lef{\`e}vre}, F. 2022, Climate of the Past, 18, 2421,
  \dodoi{10.5194/cp-18-2421-2022}

\bibitem[{{Yates} {et~al.}(2020){Yates}, {Palmer}, {Manners}, {Boutle},
  {Kohary}, {Mayne}, \& {Abraham}}]{2020MNRAS.492.1691Y}
{Yates}, J.~S., {Palmer}, P.~I., {Manners}, J., {et~al.} 2020, \mnras, 492,
  1691, \dodoi{10.1093/mnras/stz3520}

\bibitem[{{Youngblood} {et~al.}(2016){Youngblood}, {France}, {Loyd}, {Linsky},
  {Redfield}, {Schneider}, {Wood}, {Brown}, {Froning}, {Miguel}, {Rugheimer},
  \& {Walkowicz}}]{2016ApJ...824..101Y}
{Youngblood}, A., {France}, K., {Loyd}, R.~O.~P., {et~al.} 2016, \apj, 824,
  101, \dodoi{10.3847/0004-637X/824/2/101}

\bibitem[{{Zahnle} {et~al.}(2013){Zahnle}, {Catling}, \&
  {Claire}}]{2013ChGeo.362...26Z}
{Zahnle}, K.~J., {Catling}, D.~C., \& {Claire}, M.~W. 2013, Chemical Geology,
  362, 26, \dodoi{10.1016/j.chemgeo.2013.08.004}

\bibitem[{{Zhao} {et~al.}(2021){Zhao}, {Liu}, {Li}, {Liu}, \&
  {Man}}]{2021ApJ...910L...8Z}
{Zhao}, Z., {Liu}, Y., {Li}, W., {Liu}, H., \& {Man}, K. 2021, \apjl, 910, L8,
  \dodoi{10.3847/2041-8213/abebe6}

\bibitem[{Zhen {et~al.}(2022)Zhen, van Iersel, \&
  Bugbee}]{zhen2022photosynthesis}
Zhen, S., van Iersel, M.~W., \& Bugbee, B. 2022, New Phytologist, 236, 538

\bibitem[{{Zieba} {et~al.}(2023){Zieba}, {Kreidberg}, {Ducrot}, {Gillon},
  {Morley}, {Schaefer}, {Tamburo}, {Koll}, {Lyu}, {Acu{\~n}a}, {Agol}, {Iyer},
  {Hu}, {Lincowski}, {Meadows}, {Selsis}, {Bolmont}, {Mandell}, \&
  {Suissa}}]{2023arXiv230610150Z}
{Zieba}, S., {Kreidberg}, L., {Ducrot}, E., {et~al.} 2023, arXiv e-prints,
  arXiv:2306.10150, \dodoi{10.48550/arXiv.2306.10150}

\end{thebibliography}
\bibliographystyle{aasjournal}

\end{document}